\documentclass[a4paper,fleqn,usenatbib]{mnras}

\usepackage{aas_macros}
\usepackage{ae,aecompl}
\usepackage{amsmath}
\usepackage{amssymb}
\usepackage{booktabs}
\usepackage{color}
\usepackage{colortbl}
\usepackage[english]{babel}
\usepackage{enumitem}
\usepackage{epsfig}
\usepackage{fontaxes}
\usepackage[T1]{fontenc}
\usepackage{gensymb}
\usepackage{graphicx}
\usepackage{hyperref}
\usepackage{ifthen}
\usepackage{latexsym}
\usepackage{lscape}
\usepackage{mathptmx}
\usepackage{pdflscape}  
\usepackage{rotating}
\usepackage{subfig}
\usepackage{txfonts}

\title[SMBH accretion in radio-selected AGN]{SMBH accretion properties of radio-selected AGN out to $z\sim 4$}

\author[I. Delvecchio et al.]{
I.~Delvecchio,$^{1}$\thanks{E-mail: ivand@phy.hr }
V.~Smol\v{c}i\'c,$^{1}$
G.~Zamorani,$^{2}$
D.J.~Rosario,$^{3}$
M.~Bondi,$^{4}$
S.~Marchesi,$^{5}$
\newauthor T.~Miyaji,$^{6}$
M.~Novak,$^{7}$
M.T.~Sargent,$^{8}$
D.M.~Alexander,$^{3}$
J.~Delhaize$^{1}$
\\ \\
$^{1}$Department of Physics, Faculty of Science, University of Zagreb, Bijeni\v{c}ka cesta 32, 10002, Zagreb, Croatia. \\
$^{2}$INAF - Osservatorio di Astrofisica e Scienza dello Spazio - Bologna, Via Piero Gobetti 93/3, I-40129 Bologna, Italy. \\
$^{3}$Centre for Extragalactic Astronomy, Department of Physics, Durham University, South Road, Durham, DH1 3LE, UK. \\
$^{4}$Istituto di Radioastronomia di Bologna -- INAF, via P. Gobetti, 101, 40129 Bologna, Italy.\\
$^{5}$Department of Physics and Astronomy, Clemson University, Clemson, SC 29634, USA.\\
$^{6}$IAUNAM-E (Instituto de Astronom\'ia de la Universidad Nacional Aut \'onoma de M\'exico Ensenada), Ensenada, Apdo. Postal 106, Ensenada BC, 22800 Mexico. \\
$^{7}$Max-Planck-Institut f\"ur Astronomie, K\"onigstuhl 17, 69117 Heidelberg, Germany.\\
$^{8}$Astronomy Centre, Department of Physics and Astronomy, University of Sussex, Brighton, BN1 9QH, UK.
}

\date{Accepted 2018 September 14. Received 2018 August 21; in original form 2018 June 28}

\pubyear{201?}

\begin{document}
\label{firstpage}
\pagerange{\pageref{firstpage}--\pageref{lastpage}}
\maketitle

\begin{abstract}
Exploring how radio-emitting active galactic nuclei (AGN) behave and evolve with time is critical for understanding how AGN feedback impacts galaxy evolution. In this work, we investigate the relationship between 1.4~GHz radio continuum AGN luminosity ($L^{\rm AGN}_{\rm 1.4}$), specific black hole accretion rate (s-BHAR, defined as the accretion luminosity relative to the galaxy stellar mass) and redshift, for a luminosity-complete sample of radio-selected AGN in the VLA COSMOS 3~GHz Large Project. The sample was originally selected from radio-continuum observations at 3~GHz, and includes about 1800 radio AGN identified via ($>$2$\sigma$) radio-excess relative to the infrared-radio correlation of star-forming galaxies. We further select a subsample of over 1200 radio AGN that is complete in $L^{\rm AGN}_{\rm 1.4}$ over different redshift ranges, out to $z\sim 4$, and use X-ray stacking to calculate the average s-BHAR in each $L^{\rm AGN}_{\rm 1.4}$--$z$ bin. We find that the average s-BHAR is independent of $L^{\rm AGN}_{\rm 1.4}$, at all redshifts. However, we see a strong increase of s-BHAR with redshift, at fixed $L^{\rm AGN}_{\rm 1.4}$. This trend resembles the strong increase in the fraction of star-forming host galaxies (based on the $(NUV-r)$ / $(r-J)$ colours) with redshift, at fixed $L^{\rm AGN}_{\rm 1.4}$. A possible explanation for this similarity might imply a link between average AGN radiative power and availability of cold gas supply within the host galaxy. This study corroborates the idea that radio-selected AGN become more radiatively efficient towards earlier epochs, independently of their radio power.
\end{abstract}




\section{Introduction} \label{intro}

Several studies have highlighted that active galactic nuclei (AGN) are key ingredients to shape the properties and evolution of galaxies (see a comprehensive review by \citealt{hickox+2018}). In particular, it is widely accepted that every galaxy hosts a supermassive black hole (SMBH) at its centre, which may occasionally impact the surrounding interstellar medium, heating up the gas reservoirs within the galaxy, and possibly suppressing the formation of new stars (e.g. \citealt{sanders+1988}; \citealt{Hopkins+2008}). During these episodes of AGN-driven feedback, enormous amounts of energy can be relased in radiative and mechanical form, which are commonly referred to as ``quasar mode'' and ``jet mode'', respectively (e.g. \citealt{heckman+2014}). On the one hand, the ``quasar mode'' is commonly associated with radiatively-efficient ($>$1 per cent Eddington) accretion, that is fuelled by cold gas inflows onto a standard, geometrically thin and optically thick accretion disc (\citealt{shakura+1973}), and over relatively short timescales ($<$100~Myr). On the other hand, the ``jet-mode'' phase is characterized by radiatively-inefficient ($<<$1 per cent Eddington, also named ``advection-dominated accretion flow'', ADAF; e.g. \citealt{narayan+1994}), long lasting ($\sim$Gyr timescales) accretion episodes, likely fuelled via hot gas, that might generate powerful jets extending out to $>$kpc scales (e.g. \citealt{bower+2006}; \citealt{hardcastle+2007}; \citealt{Best+2012}).

Numerical simulations suggest that these two phases occur over different timescales (Myr vs. Gyr), spatial scales (pc vs. $>$kpc) and possibly over distinct evolutionary stages of the AGN-galaxy lifecycle (e.g. \citealt{Fanidakis+2012}). Nonetheless, both stages are postulated in semi-analytical models (e.g. \citealt{Croton+2016}) in order to regulate the galaxy growth, and to reproduce the colours, stellar mass and demography of today's massive galaxies (e.g. \citealt{Benson+2003}; \citealt{Hopkins+2008}).

Observational studies attempting to test the above scenario have collected large samples of AGN across different wavelengths. While optical, mid-infrared (MIR) and X-ray selected AGN are sensitive to relatively high Eddington ratios ($\lambda_{\rm edd} \gtrsim1$~per cent), radio observations can trace the non-thermal synchrotron emission powered by AGN-driven jets in the radiatively-inefficient regime. Therefore, multi-wavelength diagnostics are paramount to investigating the physical nature of each source.

In their comprehensive study, \citet{Hickox+2009} selected AGN separately at X-ray, radio and MIR wavelengths in the Bo\"otes field out to $z\sim0.8$, and analysed their host-galaxy properties. They found that the relative overlap between X-ray, radio and MIR-selected AGN is typically small ($<$10 per cent). The fractional overlap depends on the relative depth of the matched samples, as well as on the reliability and completeness of each AGN selection criterion (e.g. \citealt{Mendez+2013}). In addition, a small relative overlap may imply that different criteria are sensitive to intrinsically distinct SMBH and galaxy populations. 

This was further investigated by \citet{Goulding+2014} out to $z\sim1.4$, who confirmed that radio-selected AGN with 1.4~GHz luminosity $L_{\rm 1.4} >$~10$^{24.8}$~W~Hz$^{-1}$ are distinguishable from the other AGN classes concerning the properties of their hosts. Indeed, galaxies hosting radio AGN were found to be systematically more massive and less star forming than those hosting X-ray and MIR AGN. These and other studies (\citealt{Mendez+2013}) support the idea that the physical mechanisms driving the triggering of an AGN might be linked to the properties of its host, especially the availability of cold gas supply, that is essential to fuel both star formation and radiatively-efficient accretion onto the central SMBH \citep{Aird+2018}. However, those panchromatic studies selected relatively bright radio AGN via a conservative cut in $L_{\rm 1.4}$, thus potentially missing a large fraction of lower luminosity radio AGN, that might display a different behaviour. In this respect, deep radio surveys allow us to explore the faint (sub-mJy) radio source population with unprecedented detail, providing valuable insights into the origin of radio emission.

In the Extended Chandra Deep Field South (E-CDFS), deep observations were carried out with the Very Large Array (VLA) at 1.4~GHz down to 37$\mu$Jy~beam$^{-1}$ (at 5$\sigma$, \citealt{Miller+2013}). Exploiting these data, \citeauthor{Bonzini+2013}(\citeyear{Bonzini+2013}, \citeyear{bonzini+2015}) carried out a detailed analysis of the multi-wavelength properties of radio sources in the sub-mJy regime. They found that below 200~$\mu$Jy at 1.4~GHz, the radio source counts are dominated by normal (i.e. non-AGN) galaxies, whose radio emission originates from the diffusion of cosmic ray electrons produced in Supernova remnants within young star-forming regions. 

This aspect has been investigated further with the VLA-COSMOS~3~GHz~Large~Project \citep{Smolcic+2017a}, in which unprecedently deep radio 3~GHz observations were carried out across the full COSMOS field \citep{Scoville+2007} down to $\sim$11.5~$\mu$Jy~beam$^{-1}$ (5$\sigma$ limit). This survey allowed us to put more stringent constraints on the composition of the faint radio source population, thus confirming a change in the dominant population at 1.4~GHz flux density of 200~$\mu$Jy (e.g. \citealt{Padovani+2015}). While star-forming galaxies dominate at fainter fluxes, radio-emitting AGN take over at higher flux density levels \citep{Smolcic+2017b}. These findings do not necessarily rule out the presence of widespread AGN emission below 200~$\mu$Jy (see also \citealt{Molnar+2018}), but mainly identify the dominant source of radio emission. Pushing this issue towards even fainter radio fluxes requires high angular resolution radio observations, in order to pinpoint the circumnuclear radio AGN emission. This approach has already proved to be successful by exploiting recent data from the Very Long Baseline Interferometry (VLBI) in the COSMOS field \citep{HerreraRuiz+2017}, and will definitely become mainstream in the next decade with the Square Kilometer Array (SKA).

 However, it is well established that the radio source population above 200~$\mu$Jy consists of composite radio sources, whose 1.4~GHz emission arises from both star formation and SMBH accretion \citep{Smolcic+2017b}. The fractional contribution due to AGN activity increases towards brighter radio fluxes, however at least two radio AGN populations have been identified at these flux densities (e.g. see review by \citealt{Padovani2016}). Historically, radio AGN have been classified as ``Radio Quiet'' (RQ) or ``Radio Loud'' (RL) AGN, based on the dominant source of radio emission, namely star formation in RQ and AGN activity in RL sources, respectively. This classification has been widely used in the literature, often adopting alternative nomenclatures or selection criteria (\citealt{Smolcic+2009}; \citealt{Best+2012}; \citealt{Padovani2016}; \citealt{Mancuso+2017}), which eventually boil down to whether an excess in radio emission is significant relative to the one expected from star formation within the host. In such case, ``radio-excess AGN'' (e.g. \citealt{DelMoro+2013}; \citealt{Delvecchio+2017}, D17 hereafter) display radio emission that is mainly driven by active jets, and may be used as indicators of the AGN kinetic power (e.g. \citealt{Willott+1999}; \citealt{Cavagnolo+2010}). 

Nevertheless, radio-excess AGN might display quite heterogeneous SMBH accretion properties. Several studies focusing on the radio emission of AGN-dominated sources pointed out that these objects can span a wide range of SMBH accretion rates (e.g. \citealt{Padovani+2015}), gradually switching their dominant accretion mode between radiatively efficient and inefficient depending on $\lambda_{\rm edd}$. On the one hand, if $\lambda_{\rm edd} >$ few per cent, the radio jet power is driven by a radiatively-efficient accretion disc, possibly shining in X-ray and MIR wavelengths, and with intense optical emission lines (e.g. \citealt{Best+2012}). On the other hand, if $\lambda_{\rm edd} <<$ few per cent, the gas fuelling onto the central SMBH is radiatively inefficient, therefore the AGN bolometric output comes predominantly in kinetic form.

The connection between SMBH accretion and jet production in AGN has been largely investigated in optically-selected quasars (e.g. \citealt{Kellermann+1989}; \citealt{Cirasuolo+2003}; \citealt{Balokovic+2012}), taken from the Sloan Digital Sky Survey (SDSS) and cross-matched with FIRST data at 1.4~GHz. Although the existence of two physically distinct accretion modes in radio AGN is still debated, a typical fraction as low as $\sim$10 per cent of optically-selected quasars were detected by FIRST, suggesting a broad underlying radio power distribution. 

However, no systematic analysis of the SMBH accretion rate of radio AGN has been carried out at relatively faint radio fluxes (sub-mJy) and beyond the local Universe. Exploring these aspects down to fainter sources and towards higher redshift is crucial to answer the key question: ``Does the SMBH accretion rate of radio AGN depend on radio power and cosmic time?'' Addressing this issue requires a complete, statistical sample of radio-selected AGN spanning a wide redshift and luminosity range, as provided by the VLA-COSMOS~3~GHz~Large~Project \citep{Smolcic+2017a}.

In this paper, we explore the average SMBH accretion rate of radio-selected AGN in the COSMOS field, by exploiting the latest \textit{Chandra} data as a function of both radio power and redshift. Our sample is one of the deepest and most complete data sets of radio-excess AGN available to-date. 

The paper is structured as follows. In Sect. \ref{sample} we describe our sample selection and the identification of radio-excess AGN. The full analysis of X-ray imaging, and the average emission obtained from \textit{Chandra} stacking are presented in Sect. \ref{method}. The derivation of SMBH accretion rate estimates, as well as their relationship with radio luminosity and redshift is presented and discussed in Sect. \ref{results}. The interpretation of our results and the comparison with previous works in the literature is given in Sect. \ref{discussion}. Finally, we outline our concluding remarks in Sect. \ref{conclusions}. 

Throughout this paper, we assume a \citet{Chabrier2003} initial mass function (IMF) and a flat cosmology with $\Omega_{\rm m}$ = 0.30, $\Omega_{\rm m}$ = 0.70, and $\Omega_{\rm \Lambda}$ = 70~km~s$^{-1}$~Mpc$^{-1}$ \citep{Spergel+2003}. Magnitudes are given in the AB system \citep{Oke1974}.

\section{Sample selection and classification} \label{sample}

In this Section we describe our parent sample of VLA~3~GHz sources in the COSMOS field, their multi-wavelength counterparts, and the approach used to identify radio AGN. The final sample of radio-excess AGN that will be studied in the rest of this paper is introduced in Sect. \ref{agn_final}.

\subsection{Multi-wavelength photometry and redshifts} \label{multi_info}

\subsubsection{Optical-IR counterparts}

The sample analysed in this work was originally selected using new, highly sensitive 3~GHz observations with the Karl~G.~Jansky Very Large Array (VLA) across 2.6 square degrees of the COSMOS field, namely the ``VLA-COSMOS~3~GHz~Large Project'' \citep{Smolcic+2017a}. This is currently the deepest extra-galactic radio survey ever conducted across a medium-area field like COSMOS. We detected 10830 radio sources (at $\geq$5$\sigma$) down to an average rms of about 2.3$~\mu$Jy~beam$^{-1}$, with a 0.75'' resolution element. 

The plethora of ancillary data available in the COSMOS field enabled us to cross-match the vast majority (about 90 per cent) of our 3~GHz selected sources with optical/NIR counterparts taken from the COSMOS2015 catalogue \citep{Laigle+2016}, over an effective unmasked area of 1.77 deg$^2$, which yielded a total of 7729 matches. The counterpart association method is detailed in \citet{Smolcic+2017b}. Briefly, for each radio source we searched for all potential counterparts within a radius of 0.8'', and calculated the false-match probability of each pair by accounting for the optical/IR magnitude of each counterpart candidate. We selected the most likely counterpart via a neighbour matching algorithm, combined with a threshold in the false-match probability ($<$20 per cent), in order to minimise the fraction of spurious associations ($\sim$1 per cent). We verified that over 90 per cent of matches are unique, and found within 0.4'' (i.e. half-beam size), thus ensuring a robust association.

\subsubsection{X-ray counterparts}

In addition to the optical to far-infrared (FIR) photometry taken from the COSMOS2015 catalogue, we used the most recent X-ray data to identify AGN in our sample. X-ray sources were taken from the \textit{Chandra} COSMOS-Legacy survey (\citealt{Civano+2016}, \citealt{Marchesi+2016}). The X-ray catalogue was already cross-matched to the COSMOS2015 catalogue, as detailed in \citet{Laigle+2016}, yielding 906/7729 (around 12 per cent) sources in common with our radio-selected sample. For each X-ray source, count-rates, fluxes and absorption-corrected X-ray luminosities are available \citep{Marchesi+2016} in the soft [0.5--2 keV], hard [2--10] keV, and full [0.5--10] keV bands.

\subsubsection{Redshift measurements}

A spectroscopic or a photometric redshift was assigned to each of the 7729 radio sources with a counterpart in the COSMOS2015 catalogue (see D17). Briefly, spectroscopic redshifts were taken from the COSMOS spectroscopy master catalogue (Mara Salvato, priv. comm.). If a spectroscopic redshift was not available or of poor quality, a photometric measurement was taken instead from the COSMOS2015 catalogue \citep{Laigle+2016}, which was derived using the {\sc Le Phare} SED-fitting code (\citealt{Arnouts+1999}; \citealt{Ilbert+2006}) and employing galaxy templates from the library of \citealt{Bruzual+2003}. Special care was taken for the computation of photometric redshifts of X-ray detected sources \citep{Marchesi+2016}, that were derived by adopting a hybrid library of AGN and galaxy templates, which is more suitable for possible AGN-dominated sources (\citeauthor{Salvato+2009} \citeyear{Salvato+2009}, \citeyear{Salvato+2011}). For X-ray detected sources, we preferred this latter approach rather than relying on galaxy templates only.

Spectroscopic redshifts were collected for 2734/7729 sources (around 35 per cent), which we used to test the accuracy of the photometric redshifts. We found an excellent agreement, with a median absolute deviation $\langle |\Delta z/(1 + z)| \rangle$ = 0.010, which increases to 0.035 at $z>3$ (see also \citealt{Laigle+2016}). The full list of 7729 radio sources containing redshift and multi-wavelength information is publicly retrievable on the IPAC/IRSA database.\footnote{\url{http://irsa.ipac.caltech.edu/data/COSMOS/tables/vla/}}

\subsection{Radio-excess AGN identification} \label{agn_class}

In this Section we describe the procedure used to identify radio-excess AGN in our sample.

The full sample of 7729 radio sources was analysed with the spectral energy distribution (SED) fitting tool {\sc SED3FIT} \citep{Berta+2013}, which decomposes the broad-band SED using galaxy templates from {\sc MAGPHYS} \citep{daCunha+2008}, with the addition of AGN templates (\citealt{Fritz+2006}; \citealt{Feltre+2012}). This approach allowed us to derive AGN-corrected galaxy parameters, such as infrared luminosity ($L_{\rm IR}$), star formation rate (SFR) and stellar mass (M$_{\star}$) for each radio source.

As radio emission may arise from processes related to both star formation and AGN activity, we used the infrared-radio correlation (IRRC) to statistically decompose the total radio emission, and isolate the AGN-related radio contribution. The IRRC is often expressed in terms of the parameter $q_{\rm SF}$, defined as the logarithmic ratio between the SF-related IR luminosity ($L_{\rm IR}^{\rm SF}$) and the total 1.4~GHz radio luminosity ($L_{\rm 1.4}$). In particular, we followed the latest IRRC derivation of \citet{Delhaize+2017}, who exploited the most recent radio data from the VLA-COSMOS~3~GHz~Large~Project \citep{Smolcic+2017a} and FIR data from the \textit{Herschel Space Observatory} \citep{Pilbratt+2010}. {\it Herschel} photometry was taken with the {\it Photoconductor Array Camera and Spectrometer} (PACS; 100 and 160~$\mu$m, \citealt{Poglitsch+2010}) as part of the {\it PACS Evolutionary Probe} (PEP; \citealt{Lutz+2011}), and with the {\it Spectral and Photometric Imaging Receiver} (SPIRE; 250, 350, and 500~$\mu$m, \citealt{Griffin+2010}) from the {\it Herschel Multi-tiered Extragalactic Survey} (HerMES; \citealt{Oliver+2012}). {\it Herschel} fluxes were extracted and de-blended via a PSF fitting algorithm using {\it Spitzer} 24~$\mu$m positional priors.

Each $L_{\rm IR}^{\rm SF}$ estimate was calculated from the optical to FIR SED-fitting decomposition (see D17), after disentangling the AGN-related emission from the host-galaxy light in the total IR (rest 8-1000~$\mu$m) regime. On the other hand, $L_{\rm 1.4}$ was calculated from the integrated 3~GHz flux and $K$-corrected to 1.4~GHz by assuming a power-law radio spectrum of the shape $S_{\nu}~\propto~\nu^{\alpha}$. The spectral slope $\alpha$ was taken from the observed 1.4~GHz flux \citep{Schinnerer+2010} if the source was detected at 1.4~GHz; otherwise it was imposed to $\alpha$=--0.7 (e.g. \citealt{Condon1992}; \citealt{Murphy+2009}), consistently with the average slope measured for our 3~GHz sources \citep{Delhaize+2017}. 

Given that the infrared emission was already corrected for a possible AGN contribution, the IRRC can be used as a benchmark to quantify the excess in radio emission as a tracer of radio AGN activity. In fact, the larger the offset from $q_{\rm SF}$, the larger the AGN contribution in radio.
The above IRRC presented in \citet{Delhaize+2017} was calibrated on a sample of star-forming galaxies, after disregarding X-ray and mid-infrared AGN (see their Sect. 4.3), while still incorporating possible AGN-dominated sources in radio. While their inclusion does not impact the overall redshift trend of $q_{\rm SF}$ for star-forming galaxies \citep{Delhaize+2017}, this might drag the overall $q_{\rm SF}$--$z$ trend down, thus washing out a possible residual AGN contribution in the radio band, that might be instead relevant for our analysis. Therefore, we re-calculated the IRRC after removing the $>$2$\sigma$ outliers from either sides of the cumulative $q_{\rm SF}$ distribution, using a double-censored survival analysis to account for upper and lower limits.

This approach allowed us to partly purify the sample of star-forming galaxies from possible radio AGN contribution, leading to the following best-fit trend: $q_{\rm SF} =  (2.80 \pm 0.02) \cdot (1+z)^{-0.12 \pm 0.01}$. This relation is slightly flatter than that originally presented in \citet{Delhaize+2017}, but largely consistent with independent derivations from the literature (e.g. \citealt{Ivison+2010}; \citealt{Sargent+2010}; \citealt{Magnelli+2015}; \citealt{CalistroRivera+2017}).

This expression was then used to identify ``radio-excess AGN'': for consistency, a source was defined as ``radio-excess AGN'' (or REx AGN) if displaying a $>$2$\sigma$ offset from the above IRRC, where $\sigma$ is the observed dispersion around best-fit IRRC ($\sim$0.35~dex, consistent with the dispersion found by \citealt{Delhaize+2017}). This threshold ensures low contamination from star-forming galaxies (2.5 per cent), and it is highly sensitive to AGN-dominated ($>$80 per cent of the total 1.4~GHz emission) sources in the radio band. For each radio-excess AGN, the SF-related radio emission was quantified from the above $q_{\rm SF}(z)$ trend, and subtracted from the observed $L_{\rm 1.4}$, leaving us with the AGN-related radio luminosity at 1.4~GHz ($L^{\rm AGN}_{\rm 1.4}$). We note that our results would not change significantly if adopting the original \citet{Delhaize+2017} relation instead; however our purified $q_{\rm SF}(z)$ trend is more consistent with the adopted definition of radio-excess AGN.

\begin{figure}
     \includegraphics[width=\linewidth]{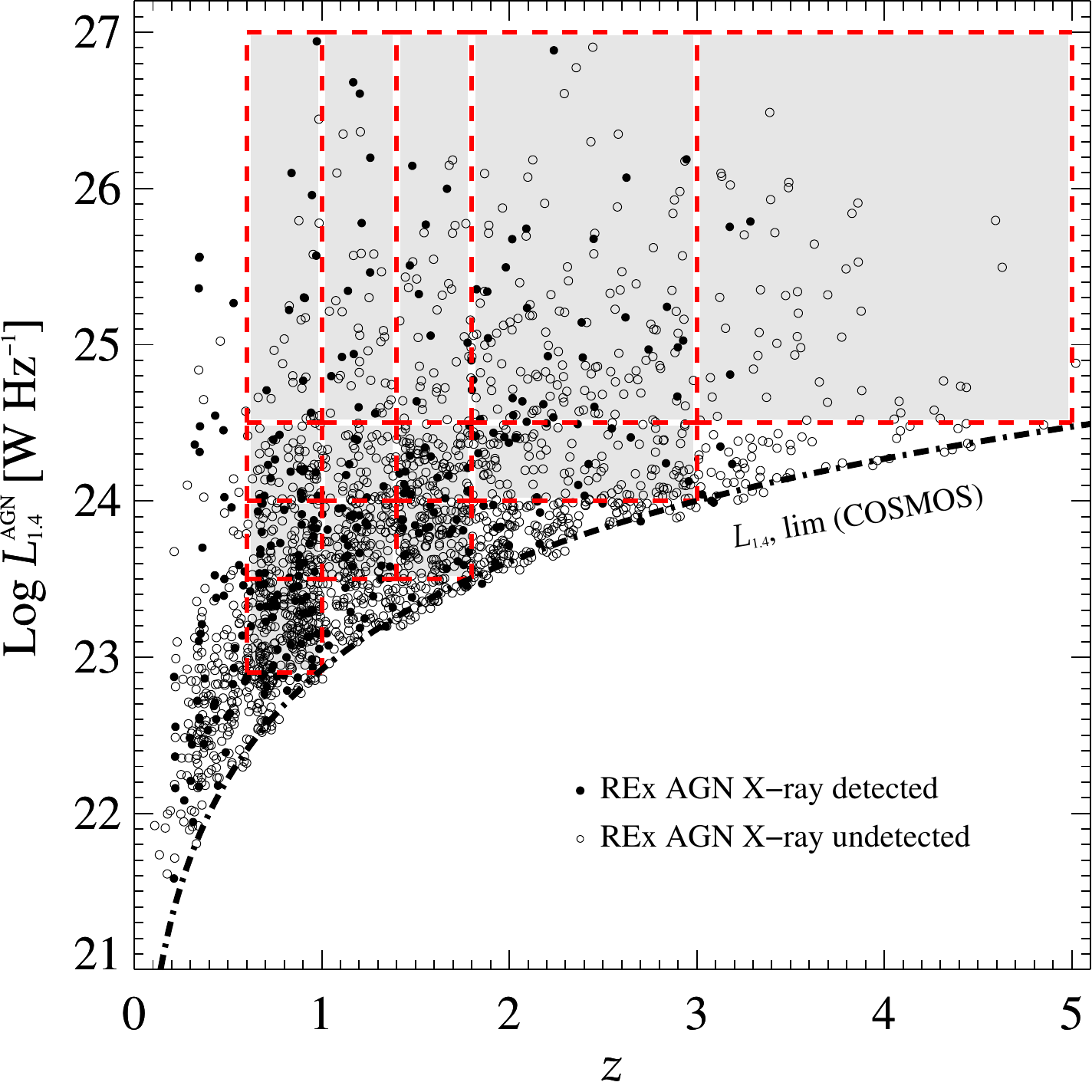}
 \caption{\small AGN-related 1.4~GHz radio luminosity ($L^{\rm AGN}_{\rm 1.4}$) as a function of redshift. The black dashed line marks the 5$\sigma$ luminosity limit of the survey, scaled to 1.4~GHz by assuming a spectral index $\alpha$=--0.7. Black circles indicate radio-excess AGN, both X-ray detected (filled circles) and undetected (empty circles). Shaded areas indicate the 13 complete $L^{\rm AGN}_{\rm 1.4}$--$z$ bins analysed in this work. }
   \label{fig:l14_z}
\end{figure}

\subsection{Final sample of radio-excess AGN} \label{agn_final}

The sample of radio-excess AGN defined in the previous Section includes 1811 sources in total. Fig. \ref{fig:l14_z} shows $L^{\rm AGN}_{\rm 1.4}$ of each source as a function of redshift. The black dashed line marks the 5$\sigma$ luminosity limit of our survey at 3~GHz, converted to 1.4~GHz by assuming a spectral index $\alpha$=--0.7 (see Sect. \ref{agn_class}). Black circles are shown for both X-ray detected (filled circles) and X-ray undetected (empty circles) radio-excess AGN. The fraction of radio AGN that are X-ray detected is only 16 per cent (286/1811) down to an average flux limit of $\sim 2\times10^{-15}$~erg~s$^{-1}$~cm$^{-2}$ in the [0.5--8] keV band. Our sample of radio-excess AGN is highly conservative (see Sect. \ref{agn_class}), while spans five orders of magnitude in $L^{\rm AGN}_{\rm 1.4}$ over the redshift range $0 <z\lesssim 5$. This sample is notably deeper in 1.4~GHz luminosity than other works identifying AGN-dominated radio sources (e.g. \citealt{Hickox+2009}; \citealt{Padovani+2015}), which allows us to investigate SMBH accretion across a wide and well-defined radio source population at different cosmic epochs.

We aim at investigating the average SMBH accretion properties of radio-excess AGN by exploiting \textit{Chandra} imaging, as a function of both $L^{\rm AGN}_{\rm 1.4}$ and redshift. However, our flux-limited radio sample is increasingly incomplete towards the faintest radio sources, at each redshift. If the X-ray properties of detected and undetected radio sources were intrinsically different, using an incomplete sample would likely bias our results. For this reason, we decided to further reduce our sample to a subsample of radio-excess AGN that is complete in $L^{\rm AGN}_{\rm 1.4}$ up to a given redshift. Fig. \ref{fig:l14_z} shows the 13 bins identified in the $L^{\rm AGN}_{\rm 1.4}$--$z$ space (shaded areas). This approach allows us to explore the simultaneous dependence of the SMBH accretion rate on both $L^{\rm AGN}_{\rm 1.4}$ and redshift, which is crucial for understanding which parameter primarily drives the accretion power in radio AGN. This cut reduces our previous radio AGN sample to 1272 sources, out of which 213 (around 17 per cent) are X-ray detected. Table \ref{tab:bins} summarizes the number of sources contained in each of the 13 $L^{\rm AGN}_{\rm 1.4}$--$z$ bins, both detected and undetected in the X-rays.
Our redshift grid at $0.6 <z< 5.0$ is optimized to ensure a good sampling of the highest radio luminosity sources, which are relatively uncommon within the volume covered by our survey at $z<0.6$. Above this redshift, we note that all X-ray detected sources have $L_{\rm X}>$10$^{42}$~erg~s$^{-1}$, which implies they are likely to be X-ray AGN.

Despite this cut, our sample remains to-date one of the largest and deepest data sets of well-defined radio-excess AGN with full X-ray coverage and redshift information available.

\begin{table}
\centering
   \caption{Number of radio-excess AGN within the 13 complete $L^{\rm AGN}_{\rm 1.4}$--$z$ bins. The number in brackets indicate the subset of X-ray detected ($d$) and not detected ($u$) sources, respectively.}
\begin{tabular}{lcc}
\hline
$ ~~~~~ z$--bin         &   $\log (L^{\rm AGN}_{\rm 1.4}$) [W~Hz$^{-1}$]    &     $\#$ (d,u)   \smallskip \\
\hline                      
$0.6  \leq z < 1.0$     &   22.9--23.5          &   232 (33, 199)   \\
                        &   23.5--24.0          &   104 (25, 79)   \\
                        &   24.0--24.5          &   67 (15, 52)   \\
                        &   24.5--27.0          &   30 (10, 20)   \smallskip \\
$1.0  \leq z < 1.4$     &   23.5--24.0          &    122 (20, 102)   \\
                        &   24.0--24.5          &    57 (6, 51)   \\
                        &   24.5--27.0          &    42 (11, 31)   \smallskip \\
$1.4  \leq z < 1.8$     &   23.5--24.0          &    140 (20, 120)   \\
                        &   24.0--24.5          &    98 (15, 83)   \\
                        &   24.5--27.0          &    46 (9, 37)   \smallskip \\
$1.8  \leq z < 3.0$     &   24.0--24.5          &    133 (16, 117)   \\
                        &   24.5--27.0          &    148 (30, 118)   \smallskip \\
$3.0  \leq z < 5.0$     &   24.5--27.0          &    53 (3, 50)   \\

\hline
\end{tabular}

\label{tab:bins}
\end{table}

\section{Calculation of the intrinsic AGN X-ray luminosity} \label{method}

In this Section we describe the method used to derive average SMBH accretion rates in each $L^{\rm AGN}_{\rm 1.4}$--$z$ bin. These were derived from the mean X-ray emission obtained from \textit{Chandra} stacks, as detailed below.

\subsection{X-ray stacking} \label{xray_stacking}

We stacked \textit{Chandra} images taken from the joint \textit{Chandra}-COSMOS and COSMOS-Legacy map (\citealt{Civano+2016}; \citealt{Marchesi+2016}), at the optical/near-infrared position of each input source, both in the soft ([0.5--2] keV), hard ([2--8] keV) and full ([0.5--8] keV) X-ray bands. 

For this task, we used the publicly available X-ray stacking tool {\sc CSTACK}\footnote{Developed by T.~Miyaji, {\sc CSTACK} is available at \url{http://lambic.astrosen.unam.mx/cstack/}}. This online software returns stacked count-rates and fluxes, as well as reliable uncertainties estimated from a bootstrapping procedure. The stacking procedure was run for each $L^{\rm AGN}_{\rm 1.4}$--$z$ bin, by combining the signal from X-ray detections and non-detections. 

We followed the same method built in {\sc CSTACK} for deriving fluxes and uncertainties from stacking. Briefly, for each $L^{\rm AGN}_{\rm 1.4}$--$z$ bin, we generated 1000 new input source lists, each created by randomly selecting the same total number of objects of the original list, but allowing duplication of the same source. We stacked each re-sampled dataset and obtained a distribution of 1000 mean (in linear scale) fluxes for each $L^{\rm AGN}_{\rm 1.4}$--$z$ bin. We considered the median value of the corresponding distribution as our best estimate of the stacked X-ray flux ($F_{\rm X}$) associated with that bin. Flux uncertainties ($dF_{\rm X}$, at 1$\sigma$ level) were estimated by interpolating the above distribution at the 16$^{\rm th}$ and 84$^{\rm th}$ percentiles. This approach enabled us to mitigate the effect of possible outliers in the underlying distribution of X-ray fluxes.

The median stacked flux derived in each $L^{\rm AGN}_{\rm 1.4}$--$z$ bin was then converted to rest-frame X-ray luminosity ($L_{\rm X}$) in each X-ray band, by assuming a power-law X-ray spectrum with an observed slope $\Gamma$=1.4, consistently with the shape of the cosmic X-ray background (e.g. \citealt{Gilli+2007}). The stacking procedure yielded $F_{\rm X}$/$dF_{\rm X} > 2$ in all bins. We note that each stacked image displays a significant detection with signal-to-noise S/N$>$5, where N was measured from the background region around the centre of each stacked image. However, throughout the paper we used $dF_{\rm X}$ as a conservative estimate of the uncertainty on the stacked flux.

\begin{figure}
\begin{center}
    \includegraphics[width=\linewidth]{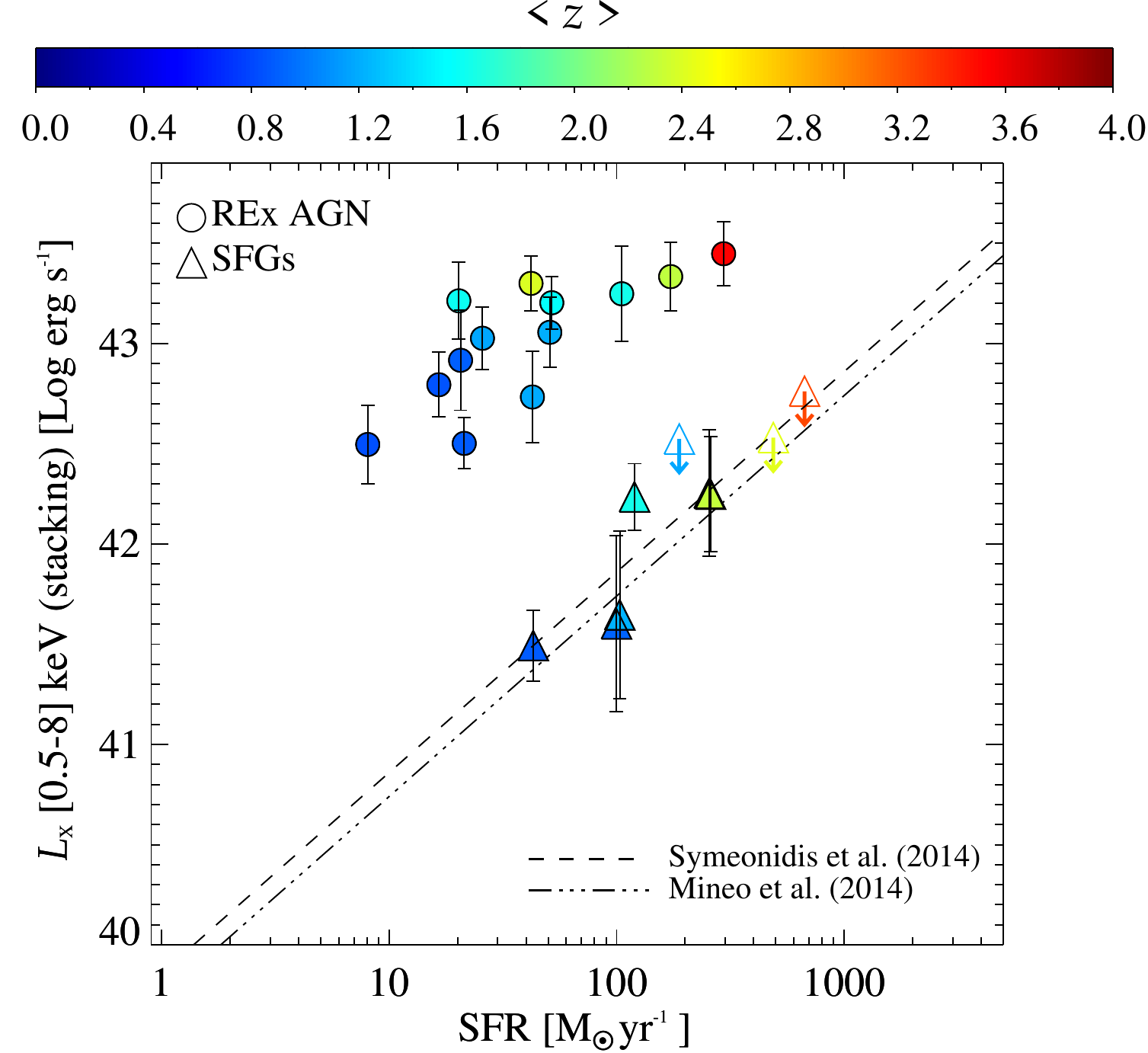}
\end{center}
 \caption{\small Comparison between the rest-frame [0.5--8] keV luminosity obtained from \textit{Chandra} stacking (colour-coded with redshift), and the (linear) mean SFR calculated from SED-fitting. The dashed line indicates the $L_{\rm X}$--SFR relation of \citet{Symeonidis+2014}, while the dotted-dashed line marks a similar relation independently proposed by \citet{Mineo+2014}. Radio-excess AGN are shown as circles, while the control sample of redshift-matched SFGs is marked with triangles. See text for details.
 }
   \label{fig:lx_sf}
\end{figure}

\subsection{Subtraction of X-ray emission due to star formation} \label{xray_sf}

We investigated the origin of the stacked X-ray emission derived in the previous Section. While X-ray emission in individually detected sources mostly arises from the AGN, for X-ray undetected sources the possible contribution of star formation might not be negligible.

We employed the empirical relation between SFR and X-ray luminosity presented in \citet{Symeonidis+2014} to estimate and then subtract the X-ray emission expected to arise from star formation. This relation was calibrated on a sample of star-forming galaxies selected with {\it Herschel} at $z<1.5$ in the {\it Chandra} Deep Field South (CDF-S), both detected and undetected in the X-rays. We estimated the SFR for each radio source from the corresponding total IR (8--1000$~\mu$m) luminosity obtained from an SED decomposition, after correcting for a possible AGN contribution (see D17).

\begin{landscape}

Fig. \ref{fig:lx_sf} illustrates the comparison between the rest-frame [0.5--8] keV luminosity obtained from stacking (see Sect. \ref{xray_stacking}) in the various redshift bins (in colour coding), and the (linear) mean SFR of the underlying radio-excess AGN population (circles). The dashed line indicates the $L_{\rm X}$--SFR relation of \citet{Symeonidis+2014}, which carries a 1$\sigma$ scatter of about a factor of two, here scaled to the [0.5--8] keV energy range. The dotted-dashed line marks a similar relation independently proposed by \citet{Mineo+2014}, shown here for comparison. 

The stacked $L_{\rm X}$ obtained in each bin displays a systematic and significant (1.2--2.2~dex) excess with respect to the X-ray emission expected from star formation, which suggests the presence of widespread AGN activity in each stacked bin. As a sanity check, we repeated the same stacking procedure for a redshift-matched control sample of (non-AGN) radio-detected star-forming galaxies (SFGs, see triangles in Fig. \ref{fig:lx_sf}), which constitute the same sample as the one previously used to calculate the $q_{\rm SF}$ parameter. For those bins in which X-ray stacking yielded no detection, we placed upper limits at the 90$^{\rm th}$ per cent confidence level (downward arrows) via bootstrapping (see Sect. \ref{xray_stacking}). Fig. \ref{fig:lx_sf} shows a very good agreement at all redshifts between the mean X-ray luminosities of our SFGs and those expected on the basis of the \citet{Symeonidis+2014} and \citet{Mineo+2014} the $L_{\rm X}$--SFR relations. This check ensures the applicability of the $L_{\rm X}$--SFR relation for our sample of radio-excess AGN and throughout the full redshift range. 

In each $L^{\rm AGN}_{\rm 1.4}$--$z$ bin, we used the \citet{Symeonidis+2014} relation to estimate and then subtract the X-ray emission expected expected to arise from star formation, at the mean SFR of the stacked sample ($<$10 per cent in all bins). Only the AGN-related X-ray emission was considered in the next steps. The mean $L_{\rm X}$ values obtained from stacking are reported in Table \ref{tab:stacking}, separately for the full radio-excess sample, and for the subset with X-ray detection. As shown in Table \ref{tab:stacking}, the X-ray detected subsample displays, on average, 0.5--0.8~dex higher X-ray emission than that of the full (i.e. X-ray detected+undetected) radio-excess AGN sample. We stress that a weaker (0.3--1.5~dex), but systematic X-ray excess would be still in place if only undetected X-ray sources were considered. This implies that, also below the \textit{Chandra} detection limit, a significant fraction of the stacked X-ray emission is arising from active SMBH accretion.

\begin{table} 

\centering
   \caption{Main parameters (and corresponding 1$\sigma$ uncertainties) obtained via X-ray stacking in each $L^{\rm AGN}_{\rm 1.4}$--$z$ bin. We calculated the mean value of each parameter, separately for all radio-excess AGN (both X-ray detected and undetected, see tag ``all''), and for the X-ray detected subset alone (see tag ``det''). The tag $ L_{\rm X}$ corresponds to the [0.5--8] keV X-ray luminosity obtained from \textit{Chandra} stacking, while the tag $ L^{\rm AGN}_{\rm X}$ represents the intrinsic AGN X-ray luminosity. }
\begin{tabular}{lccccccccccc}
\hline
$ ~~~~~ z$--bin         &   $L^{\rm AGN}_{\rm 1.4}$   &     $ L_{\rm X}$ &   $ L_{\rm X}$  &  HR  &  HR   &   $ L^{\rm AGN}_{\rm X}$ &  $ L^{\rm AGN}_{\rm X}$  &   s-BHAR & s-BHAR  &  $\log (\lambda_{\rm edd})$   &  $\log (\lambda_{\rm edd})$    \\
    &    $\log$[W~Hz$^{-1}$]   &  $\log$[erg~s$^{-1}$]  &  $\log$[erg~s$^{-1}$]  &   &    &   $\log$[erg~s$^{-1}$]  &  $\log$[erg~s$^{-1}$]  &   $\log$[erg~s$^{-1} M_{\odot}^{-1}$]  &    $\log$[erg~s$^{-1} M_{\odot}^{-1}$]  &    &      \\
    &      &   (all) &   (det) &   (all)   &   (det) &   (all)   &   (det) &    (all) &   (det) &    (all) &   (det)   \\
\hline                      
   \smallskip  \\ $0.6 \leq z < 1.0$  &  22.9$-$23.5  &  42.50${\pm 0.18}$  &  
43.25${\pm 0.18}$  &  -0.29${\pm 0.40}$  &  -0.09${\pm 0.12}$  &  
42.61${\pm 0.21}$  &  43.44${\pm 0.20}$  &  32.35${\pm 0.20}$  &  
33.25${\pm 0.23}$  &  -3.06${\pm 0.20}$  &  -2.15${\pm 0.23}$  \\ 
  \smallskip 
  &  23.5$-$24.0  &  42.79${\pm 0.15}$  &  43.41${\pm 0.13}$  &  
0.04${\pm 0.28}$  &  0.09${\pm 0.11}$  &  43.01${\pm 0.17}$  &  
43.64${\pm 0.15}$  &  32.71${\pm 0.18}$  &  33.35${\pm 0.18}$  &  
-2.69${\pm 0.18}$  &  -2.05${\pm 0.18}$  \\ 
  \smallskip 
  &  24.0$-$24.5  &  42.50${\pm 0.11}$  &  43.03${\pm 0.11}$  &  
0.05${\pm 0.49}$  &  0.08${\pm 0.17}$  &  42.71${\pm 0.13}$  &  
43.25${\pm 0.12}$  &  32.35${\pm 0.13}$  &  33.04${\pm 0.13}$  &  
-3.05${\pm 0.13}$  &  -2.36${\pm 0.13}$  \\ 
  \smallskip 
  &  24.5$-$27.0  &  42.92${\pm 0.23}$  &  43.46${\pm 0.18}$  &  
-0.15${\pm 0.25}$  &  -0.15${\pm 0.12}$  &  43.08${\pm 0.27}$  &  
43.63${\pm 0.22}$  &  32.69${\pm 0.29}$  &  33.30${\pm 0.24}$  &  
-2.71${\pm 0.29}$  &  -2.10${\pm 0.24}$  \\ 
  \smallskip 
  \smallskip  \\ $1.0 \leq z < 1.4$  &  23.5$-$24.0  &  43.03${\pm 0.14}$  &  
43.65${\pm 0.15}$  &  0.06${\pm 0.30}$  &  0.05${\pm 0.11}$  &  
43.25${\pm 0.17}$  &  43.87${\pm 0.16}$  &  32.96${\pm 0.19}$  &  
33.62${\pm 0.18}$  &  -2.44${\pm 0.19}$  &  -1.79${\pm 0.18}$  \\ 
  \smallskip 
  &  24.0$-$24.5  &  42.73${\pm 0.21}$  &  43.58${\pm 0.16}$  &  
-0.13${\pm 0.47}$  &  -0.06${\pm 0.12}$  &  42.89${\pm 0.23}$  &  
43.78${\pm 0.18}$  &  32.55${\pm 0.24}$  &  33.62${\pm 0.22}$  &  
-2.85${\pm 0.24}$  &  -1.78${\pm 0.22}$  \\ 
  \smallskip 
  &  24.5$-$27.0  &  43.06${\pm 0.15}$  &  43.61${\pm 0.13}$  &  
-0.10${\pm 0.31}$  &  -0.05${\pm 0.13}$  &  43.23${\pm 0.17}$  &  
43.81${\pm 0.14}$  &  32.84${\pm 0.19}$  &  33.51${\pm 0.16}$  &  
-2.56${\pm 0.19}$  &  -1.89${\pm 0.16}$  \\ 
  \smallskip 
  \smallskip  \\ $1.4 \leq z < 1.8$  &  23.5$-$24.0  &  43.21${\pm 0.18}$  &  
43.97${\pm 0.18}$  &  -0.12${\pm 0.36}$  &  -0.12${\pm 0.11}$  &  
43.39${\pm 0.20}$  &  44.15${\pm 0.21}$  &  33.36${\pm 0.23}$  &  
34.15${\pm 0.25}$  &  -2.04${\pm 0.23}$  &  -1.25${\pm 0.25}$  \\ 
  \smallskip 
  &  24.0$-$24.5  &  43.20${\pm 0.12}$  &  43.77${\pm 0.09}$  &  
-0.00${\pm 0.36}$  &  0.01${\pm 0.13}$  &  43.41${\pm 0.14}$  &  
43.99${\pm 0.10}$  &  33.16${\pm 0.15}$  &  33.90${\pm 0.12}$  &  
-2.24${\pm 0.15}$  &  -1.50${\pm 0.12}$  \\ 
  \smallskip 
  &  24.5$-$27.0  &  43.25${\pm 0.20}$  &  43.89${\pm 0.20}$  &  
-0.15${\pm 0.37}$  &  -0.16${\pm 0.12}$  &  43.40${\pm 0.23}$  &  
44.05${\pm 0.22}$  &  33.12${\pm 0.27}$  &  33.84${\pm 0.27}$  &  
-2.28${\pm 0.27}$  &  -1.56${\pm 0.27}$  \\ 
  \smallskip 
  \smallskip  \\ $1.8 \leq z < 3.0$  &  24.0$-$24.5  &  43.30${\pm 0.13}$  &  
44.15${\pm 0.11}$  &  0.02${\pm 0.50}$  &  -0.03${\pm 0.13}$  &  
43.53${\pm 0.15}$  &  44.37${\pm 0.12}$  &  33.56${\pm 0.17}$  &  
34.42${\pm 0.15}$  &  -1.84${\pm 0.17}$  &  -0.98${\pm 0.15}$  \\ 
  \smallskip 
  &  24.5$-$27.0  &  43.33${\pm 0.16}$  &  43.98${\pm 0.14}$  &  
0.04${\pm 0.50}$  &  0.06${\pm 0.17}$  &  43.55${\pm 0.18}$  &  
44.22${\pm 0.16}$  &  33.32${\pm 0.20}$  &  33.98${\pm 0.20}$  &  
-2.08${\pm 0.20}$  &  -1.42${\pm 0.20}$  \\ 
  \smallskip 
  \smallskip  \\ $3.0 \leq z < 5.0$  &  24.5$-$27.0  &  43.45${\pm 0.14}$  &  
44.26${\pm 0.07}$  &  0.14${\pm 0.42}$  &  0.28${\pm 0.13}$  &  
43.68${\pm 0.18}$  &  44.57${\pm 0.09}$  &  33.67${\pm 0.22}$  &  
34.62${\pm 0.11}$  &  -1.73${\pm 0.22}$  &  -0.78${\pm 0.11}$  \\

\hline
\end{tabular}

\label{tab:stacking}

\end{table}
  
\end{landscape}

\begin{figure*}
\begin{center}
    \includegraphics[width=\linewidth]{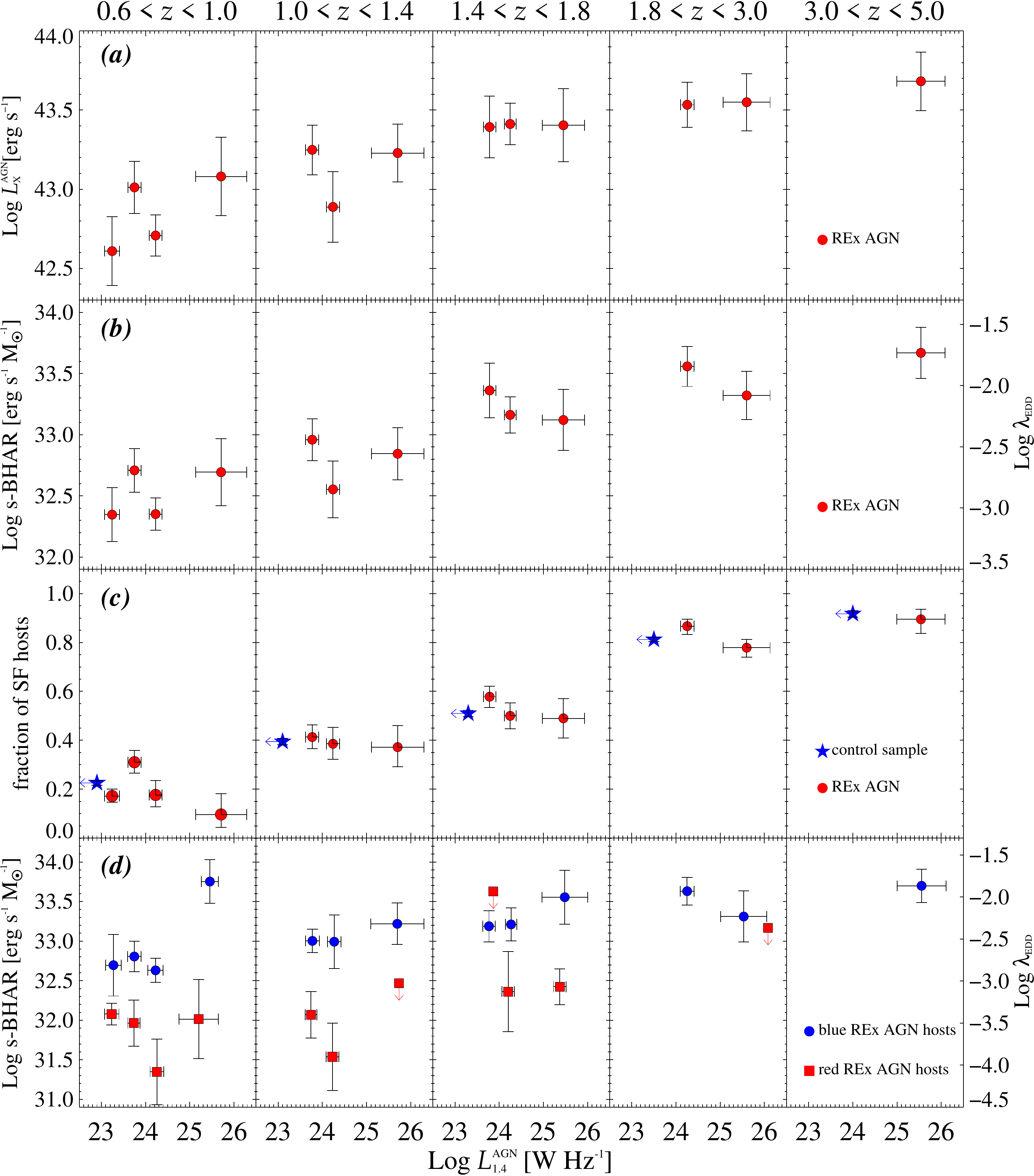}
\end{center}
 \caption{\small Average parameters (y-axis), as a function of $L^{\rm AGN}_{\rm 1.4}$ (x-axis), separately in different redshift bins. In each panel, the horizontal error bars indicate the 1$\sigma$ dispersion in $L^{\rm AGN}_{\rm 1.4}$ of the underlying sample, while the vertical error bars indicate the corresponding $\sigma$ uncertainty. The plot consists in four sets of panels. (a): Average $L^{\rm AGN}_{\rm X}$ obtained from stacking (Sect. \ref{average_lx}) after subtracting the star formation contribution (Sect. \ref{xray_sf}) and correcting for nuclear obscuration (Sect. \ref{xray_corr}). (b): Average s-BHAR (or $\lambda_{\rm edd}$) obtained after scaling $L^{\rm AGN}_{\rm X}$ to bolometric AGN luminosity, and assuming a matter-to-radiation conversion efficiency of 0.1 (Sect. \ref{average_sbhar}). (c): Fraction of (blue) star-forming radio AGN hosts, based on the $(NUV-r)$ / $(r-J)$ diagram (Sect. \ref{colours}). The $\pm$~1$\sigma$ uncertainties were derived by following \citet{Gehrels1986}. Red circles indicate radio-excess AGN, while blue stars refer to a control sample of non-AGN galaxies matched in M$_{\star}$ and redshift. The left-pointing arrows indicate the 3~GHz luminosity (5$\sigma$) limit of our survey at the mean redshift of the underlying sample (scaled to 1.4~GHz by assuming a spectral index $\alpha$=--0.7). (d): Average s-BHAR (or $\lambda_{\rm edd}$) derived separately for blue (circles) and red (squares) radio-excess AGN hosts, based on the $(NUV-r)$ / $(r-J)$ diagram (Sect. \ref{colours} and Fig. \ref{fig:nuvrj}). Down-pointing arrows indicate upper limits at 90\% confidence level. 
 }
   \label{fig:merged}
\end{figure*}
%

\subsection{Correction for nuclear obscuration} \label{xray_corr}

The AGN-related X-ray emission derived from \textit{Chandra} stacking might be underestimated due to nuclear obscuration. In this Section, we correct the stacked X-ray emission for this effect via the Hardness Ratio (HR), by following the procedure adopted by \citet{Marchesi+2016}. The HR is defined as $\rm{HR} = \frac{H - S}{H + S}$, where (H,S) represent the (exposure weighted) photon counts in the hard (2--8 keV) and soft (0.5--2 keV) band, respectively. 

A large fraction (about 85 per cent) of our input sources are individually undetected in all \textit{Chandra} bands. In the low-count regime, uncertainties do not follow a Gaussian distribution and should be treated in a more appropriate way. The \textit{Bayesian Estimation of Hardness Ratios} (BEHR) method \citep{Park+2006} is particularly effective for faint X-ray sources, because it does not need a detection in both bands to work and it runs a Bayesian Markov chain Monte Carlo calculation to estimate errors. Starting from the (H,S) values obtained from X-ray stacking in each bin, we ran BEHR recursively by comparing - at each iteration - the observed HR inferred with BEHR to an intrinsic (i.e. unabsorbed) X-ray spectral model, that was assumed to be a power-law spectrum with photon index $\Gamma$=1.8 (e.g. \citealt{Tozzi+2006}). This method allowed us to obtain a rough estimate of the average nuclear obscuration, that we used to calculate the mean absorption-corrected AGN X-ray luminosity ($L^{\rm AGN}_{\rm X}$) and its uncertainty in each $L^{\rm AGN}_{\rm 1.4}$--$z$ bin. The average obscuration correction factor to the [0.5--8] keV AGN X-ray luminosity ranges from 1.3 to 1.8, independently of $L^{\rm AGN}_{\rm 1.4}$ or redshift. The HR values calculated with BEHR are listed in Table \ref{tab:stacking}, separately for the full radio-excess AGN sample and the subset with X-ray detection. The HR values are consistent between these two populations within the 1$\sigma$ uncertainties.

We note that the level of obscuration based on the HR might be poorly constrained in the case of highly obscured AGN \citep{Xue+2011}, therefore we cross-matched our sample with the subset of 67 Compton Thick (i.e. with hydrogen column density N$_{\rm H}>$1.5$\times$10$^{24}$~cm$^{-2}$) AGN candidates identified from X-ray spectral analysis in the \textit{Chandra} Legacy survey (\citealt{Lanzuisi+2018}, see also \citealt{Marchesi+2016b}). We found only 11 common matches in total (11/67 $\sim$ 16 per cent, higher than, but consistent with the fraction of X-ray selected AGN in COSMOS that are also radio-excess AGN, 11 per cent), suggesting that the average intrinsic X-ray emission of radio-excess AGN might not arise primarily from highly obscured AGN accretion. Our typical N$_{\rm H}$ estimates span the range 10$^{22-23}$~cm$^{-2}$, suggesting moderate level of obscuration. In this N$_{\rm H}$ regime, previous studies have shown that HR-based N$_{\rm H}$ estimates agree fairly well with those based on X-ray spectra (e.g. \citealt{Xue+2011}).

\section{Results} \label{results}

We used the $L^{\rm AGN}_{\rm X}$ estimates derived in Sect. \ref{xray_corr} to study the simultaneous dependence of $L^{\rm AGN}_{\rm X}$ on $L^{\rm AGN}_{\rm 1.4}$ and redshift.

\subsection{Average $L^{\rm AGN}_{\rm X}$ of radio-excess AGN} \label{average_lx}

Fig. \ref{fig:merged} (panel {\it a}) illustrates the average absorption-corrected $L^{\rm AGN}_{\rm X}$ as a function of $L^{\rm AGN}_{\rm 1.4}$--$z$ and separately in each redshift bin. We calculated the 1$\sigma$ uncertainties on the derived $L^{\rm AGN}_{\rm X}$ (see Sect. \ref{xray_corr}) by propagating the uncertainties on both the subtraction of X-ray star formation and the correction for nuclear obscuration, via a Monte Carlo approach. These plots show a strong increase of $L^{\rm AGN}_{\rm X}$ with redshift, at fixed $L^{\rm AGN}_{\rm 1.4}$, while no clear trend appears as a function of $L^{\rm AGN}_{\rm 1.4}$, within the same redshift bin. These findings suggest that the AGN radiative power traced by X-ray emission is not tightly tied to the radio AGN power traced by our VLA data, in a statistical sense. In addition, and especially for the highest $L^{\rm AGN}_{\rm 1.4}$ range, our results highlight an enhancement of the average $L^{\rm AGN}_{\rm X}$ from $z\sim$0.7 to $z\sim$3.5 by a factor of about four. 

We remind the reader that our radio-excess AGN sample is complete in radio luminosity in each $L^{\rm AGN}_{\rm 1.4}$--$z$ bin. This ensures that our observed trends are not affected by obvious selection effects. We also stress that the stronger redshift dependence seen here is not driven by X-ray emission being boosted by a few bright detections. As shown in Table \ref{tab:bins}, the fraction of X-ray detections increases for brighter radio sources and at lower redshifts. If our $L^{\rm AGN}_{\rm X}$ estimates were biased high due to bright X-ray outliers, we would expect to see $L^{\rm AGN}_{\rm X}$ to resemble the same trend of the fraction of X-ray detections. On the contrary, we observed a flat trend between $L^{\rm AGN}_{\rm X}$ and $L^{\rm AGN}_{\rm 1.4}$, and a positive evolution of $L^{\rm AGN}_{\rm X}$ with redshift (see panel {\it(a)} of Fig. \ref{fig:merged}). This check suggests that the relative fraction of X-ray detected or undetected sources within a given bin does not significantly affect the average X-ray emission obtained from stacking.
The only mild enhancement of $L^{\rm AGN}_{\rm X}$ with $L^{\rm AGN}_{\rm 1.4}$ can be identified in the lowest redshift bin, though marginally consistent also with a flat trend. This apparent increase may be partly attributed to the relatively high fraction of X-ray detections in the highest $L^{\rm AGN}_{\rm 1.4}$ bin (33 per cent against an average 16 per cent), as shown in Table \ref{tab:bins}.

\begin{figure}
\begin{center}
    \includegraphics[width=\linewidth]{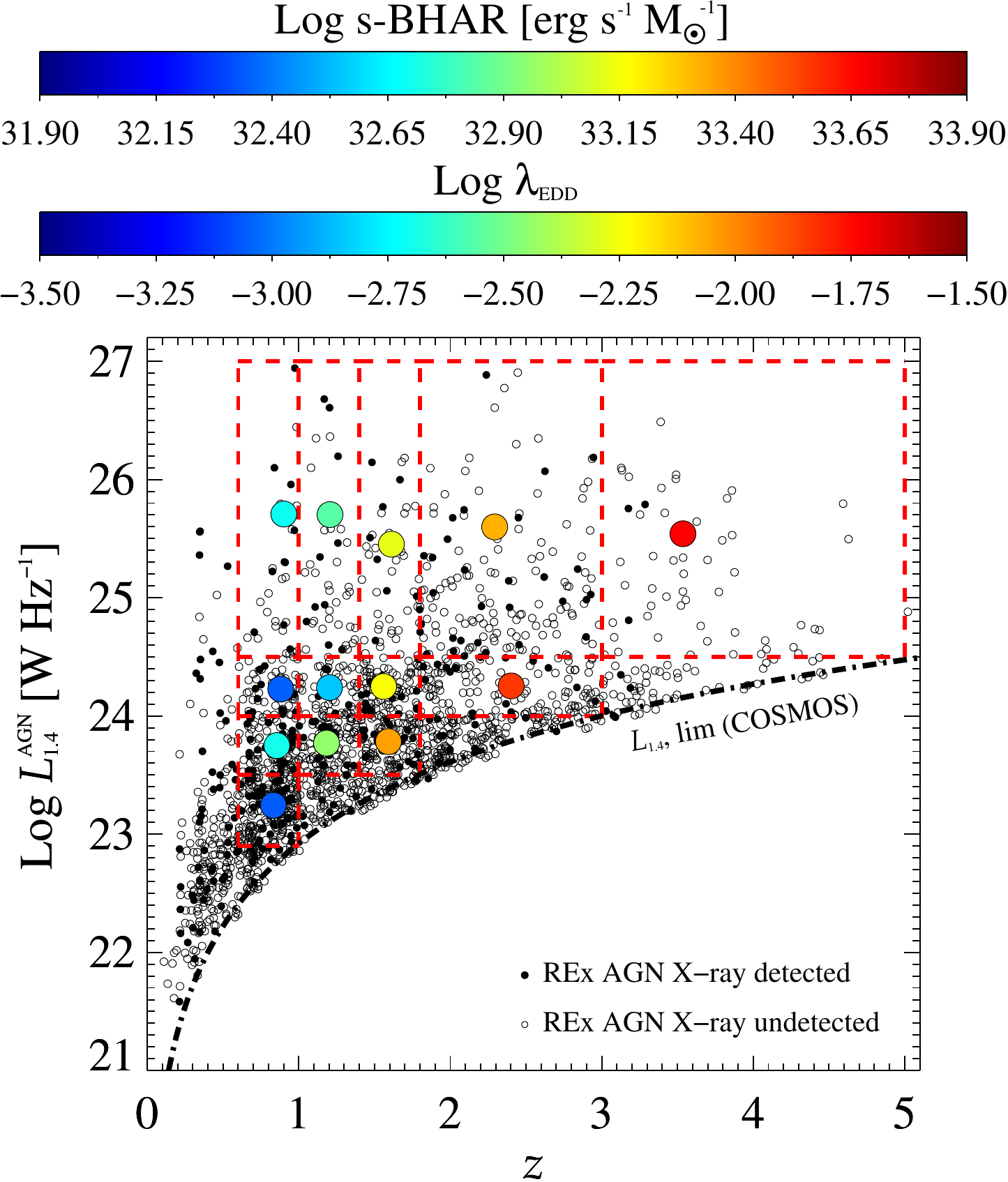}
\end{center}
 \caption{\small Average s-BHAR (or $\lambda_{\rm edd}$, in colour coding) derived in each complete $L^{\rm AGN}_{\rm 1.4}$--$z$ bin via X-ray stacking of both detected and undetected X-ray sources.
 }
   \label{fig:lr_vs_z_edd}
\end{figure}

\subsection{From $L^{\rm AGN}_{\rm X}$ to specific SMBH accretion rate} \label{average_sbhar}

We used the X-ray luminosity obtained from stacking to derive the average SMBH accretion rate, in each $L^{\rm AGN}_{\rm 1.4}$--$z$ bin. We converted the average absorption-corrected $L^{\rm AGN}_{\rm X}$ to SMBH accretion rate by using a set of scaling factors. Each intrinsic $L^{\rm AGN}_{\rm X}$ was first scaled to bolometric AGN luminosity ($L_{\rm bol}$) via a set of luminosity-dependent bolometric corrections \citep{Lusso+2012}, and then turned into SMBH accretion rate, by assuming a constant matter-to-radiation conversion efficiency of 0.1 (see review by \citealt{Alexander+2012}). Finally, this parameter was converted to $\lambda_{\rm edd}$ by assuming a constant ratio between SMBH mass (M$_{\rm BH}$) and M$_{\star}$ of 0.002 (e.g. \citealt{Haring+2004}). The latter assumption carries severe uncertainties, and it is poorly constrained for non-local AGN, especially if not type-1, unobscured systems (e.g. \citealt{Ricci+2017}). For this reason, we expressed the $\lambda_{\rm edd}$ in terms of ``specific SMBH accretion rate'' (s-BHAR), which is defined as $L_{\rm bol}$ normalized to the M$_{\star}$ of the host (see \citealt{Aird+2012}). This parameter is roughly equivalent to $\lambda_{\rm edd}$, though it does not assume a specific conversion from M$_{\rm BH}$ to M$_{\star}$.

Fig. \ref{fig:merged} (panel {\it b}) shows the average s-BHAR (or $\lambda_{\rm edd}$) as a function of $L^{\rm AGN}_{\rm 1.4}$, in each redshift bin. An equivalent plot is shown in Fig. \ref{fig:lr_vs_z_edd} with the average s-BHAR colour-coded in the $L^{\rm AGN}_{\rm 1.4}$--$z$ diagram. From these plots, the average s-BHAR appears to increase with redshift, at fixed $L^{\rm AGN}_{\rm 1.4}$, while the trend with $L^{\rm AGN}_{\rm 1.4}$ is consistent with being flat, within any given redshift bin. We note that the error bars displayed in Fig. \ref{fig:merged} reflect the same sources of uncertainties discussed in \ref{average_lx}, simply scaled to s-BHAR (or $\lambda_{\rm edd}$). Mean s-BHAR and $\lambda_{\rm edd}$ values are listed in Table \ref{tab:stacking} with their corresponding 1$\sigma$ uncertainties, for both the full radio-excess AGN sample and for the X-ray detected subsample alone.

The rising trend of s-BHAR with redshift appears stronger than that observed for the average $L^{\rm AGN}_{\rm X}$. This is partly due to the luminosity-dependent bolometric correction, but also to the SMBH accretion rate being normalized to the host galaxy M$_{\star}$. Indeed, the average galaxy M$_{\star}$ tends to decrease with increasing redshift, since galaxies at earlier cosmic epochs have not built up their full stellar content. These factors slightly amplify the previous redshift trend seen with $L^{\rm AGN}_{\rm X}$, leading to the average s-BHAR increase by a factor of $\sim$10 from $z\sim$0.7 to $z\sim$3.5, in the highest 1.4~GHz luminosity range (10$^{24.5}< L^{\rm AGN}_{\rm 1.4}<$10$^{27}$~ W~Hz$^{-1}$). While the full redshift range ($0.6 <z< 5.0$, split in five bins) can be explored for the brightest radio sources, we note that a smaller redshift window is accessible for fainter radio sources, counting four (three) redshift bins for the 1.4~GHz luminosity range 10$^{24}< L^{\rm AGN}_{\rm 1.4}<$10$^{24.5}$~ W~Hz$^{-1}$ (10$^{23.5}< L^{\rm AGN}_{\rm 1.4}<$10$^{24}$~ W~Hz$^{-1}$). However, the redshift increase is observed in all $L^{\rm AGN}_{\rm 1.4}$ bins.

In particular, at $z>2$ the average s-BHAR corresponds to $\lambda_{\rm edd} \gtrsim$1 per cent, implying that the typical radio AGN activity can be approximately described with radiatively efficient accretion (e.g. \citealt{Merloni+2008}). We acknowledge the notably high uncertainties and possible systematics of the derived $\lambda_{\rm edd}$ estimates, thus we caution the reader that different scaling factors might lead to a wide range of $\lambda_{\rm edd}$ values. Nonetheless, under the basic assumption that the above-mentioned scaling factors apply regardless of $L^{\rm AGN}_{\rm 1.4}$ and redshift, our results suggest that radio AGN with a given jet power (or $L^{\rm AGN}_{\rm 1.4}$) become more and more radiatively efficient from low to high redshift. A more detailed discussion of the implications of these findings is given in Sect. \ref{discussion}.

\begin{figure}
     \includegraphics[width=\linewidth]{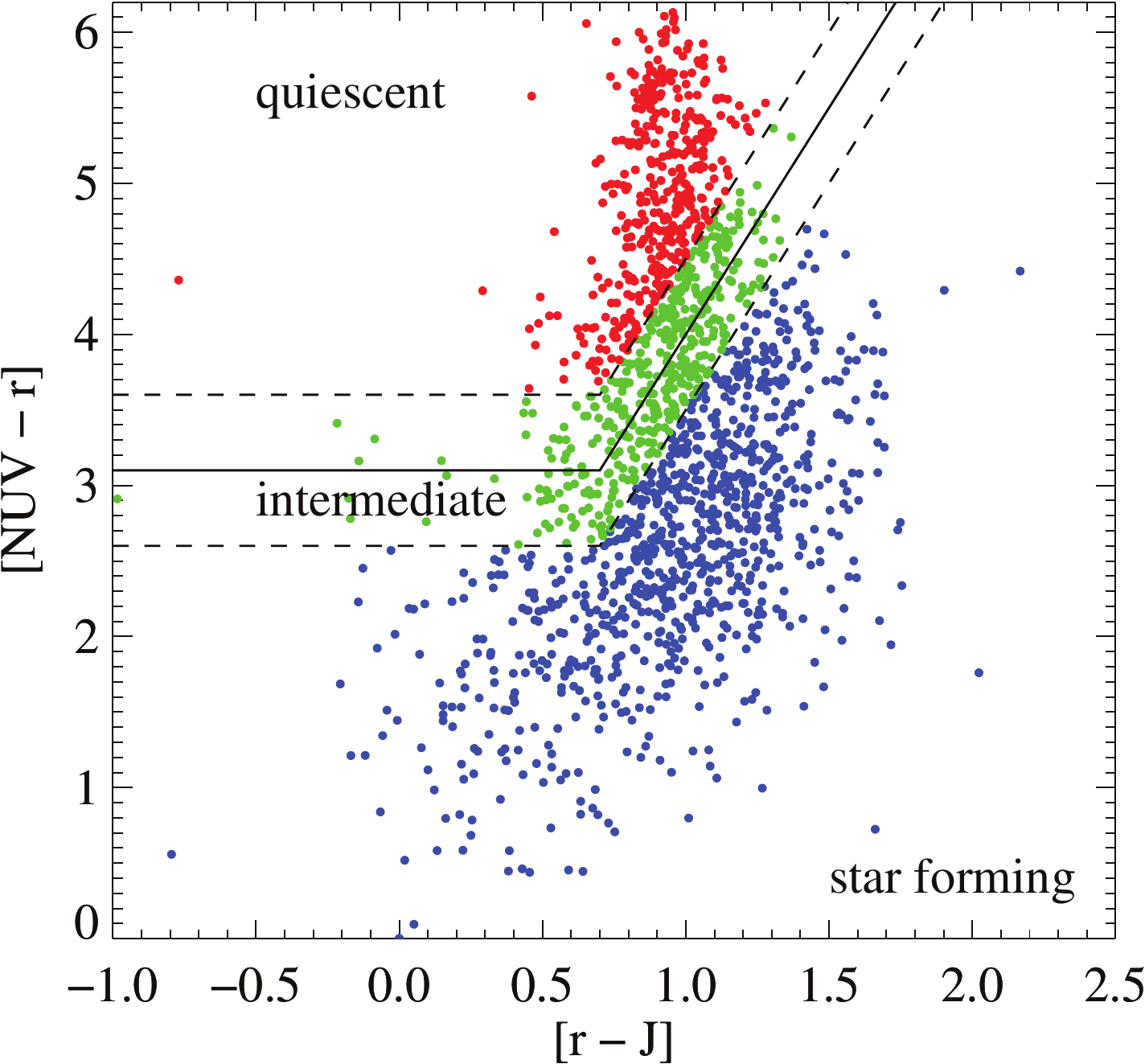}
     
 \caption{\small Rest-frame $(NUV-r)$ v $(r-J)$ diagram. The dashed lines mark the intermediate wedge centred on the solid line (Eq.\ref{eq:1}), which separates quiescent (upper-left corner) and star-forming galaxies (lower-right corner). Circles indicate the radio-excess AGN sample studied in this work, whose hosts are classified based on the $(NUV-r)/(r-J)$ as quiescent (red), intermediate (green) or star-forming (blue). See text for details.
 }
   \label{fig:nuvrj}
\end{figure}

\subsection{Colours of radio-excess AGN hosts} \label{colours}

In this Section we explore the star-forming properties of galaxies hosting radio-excess AGN. The reason for this analysis is to investigate whether the trends of s-BHAR as a function of redshift and $L^{\rm AGN}_{\rm 1.4}$ seen in Sect. \ref{average_sbhar} are linked to the integrated properties of the host galaxy. For each radio-excess AGN, we have an optical-NIR counterpart from the COSMOS2015 catalogue of \citet{Laigle+2016}. Each entry lists K-corrected absolute magnitudes in the rest-frame $NUV$, $r$ and $J$ bands, which we used to plot our galaxies in the colour-colour diagram $(NUV-r)$ / $(r-J)$. Then we used the following empirical expressions (from \citealt{Davidzon+2017}):
\begin{equation}
 (NUV - r) = 3(r - J) + 1 ~~~~~ and ~~~~~ (NUV - r) > 3.1
      \label{eq:1}
\end{equation} 
to set the dividing line between quiescent galaxies (above the locus) and SFGs (below the locus), which roughly corresponds to a cut in specific SFR (s-SFR = SFR/M$_{\star}$, \citealt{Ilbert+2013}). Our Fig.~\ref{fig:nuvrj} shows the distribution of radio-excess AGN analysed in this work (circles) over the $(NUV-r)$ / $(r-J)$ diagram. The black solid line corresponds to Eq.~\ref{eq:1}: this line does not separately classify galaxies within the so-called ``green valley'', i.e. the intermediate wedge between quiescent and star-forming systems. To account for this effect, we conservatively select as (``blue'') SFGs only those sources lying more than 0.5~mag below Eq. \ref{eq:1} (i.e. blue circles in Fig.~\ref{fig:nuvrj}). This colour-colour diagram reduces the possible mis-classification of highly dust-obscured galaxies as passive systems, and it has been proved to be effective also at $z>4$ (see Fig. 7 in \citealt{Davidzon+2017}).

Following this approach, we calculated the fraction of star-forming hosts (f$_{\rm SF}$) within our radio-excess AGN sample, as a function of $L^{\rm AGN}_{\rm 1.4}$ and redshift. Fig. \ref{fig:merged} (panel {\it c}) shows the derived f$_{\rm SF}$ estimates for radio-excess AGN (red circles), while the error bars indicate their corresponding $\pm$~1$\sigma$ confidence interval (according to Eqs. 23--24 from \citealt{Gehrels1986}). Fig. \ref{fig:merged} suggests a strong increase of f$_{\rm SF}$ with redshift, at fixed $L^{\rm AGN}_{\rm 1.4}$, at least up to $z\sim2$, while there is no, or at most a weakly decreasing trend with $L^{\rm AGN}_{\rm 1.4}$, at fixed redshift.

These observed trends are qualitatively similar to those seen for the s-BHAR (or $\lambda_{\rm edd}$) in Sect. \ref{average_sbhar}. This similarity might imply a link, in a statistical sense, between the average AGN radiative power and active star formation within radio-excess AGN hosts. A positive correlation between BHAR and SFR has been previously documented for IR-selected samples of galaxies (\citealt{Chen+2013}; \citealt{Delvecchio+2015}), and justified via numerical simulations under the assumption of a common fuelling mechanisms for both SMBH accretion and galaxy star formation (e.g. \citealt{McAlpine+2017}). However, hints of such a correlation have never been observed for radio AGN hosts, mostly because of the lack of deep and uniform radio data across a wide $L^{\rm AGN}_{\rm 1.4}$ and redshift range. 

As a sanity check, we investigated whether a similar trend could be observed for a control sample of ``inactive'' galaxies (i.e. not radio-excess AGN) matched in M$_{\star}$ and redshift. This latter sample was selected from the full COSMOS2015 catalogue \citep{Laigle+2016}, after eliminating all radio-excess AGN (from this work) and sources within masked areas in the optical/NIR images. This sample was then split in five redshift bins, as previously done for our radio-excess AGN. Within each bin, we randomly selected a subset of COSMOS2015 sources in different M$_{\star}$ bins, in order to match the M$_{\star}$ distribution observed for radio-excess AGN. We applied this method to all redshift bins, and retrieved the $(NUV-r)$ / $(r-J)$ colours for each randomly selected source. The blue stars in Fig. \ref{fig:merged} (panel {\it c}) mark the f$_{\rm SF}$ values for the above-mentioned control sample. Since this mainly consists of VLA 3~GHz undetected sources (though not necessarily, since there might be a subset of radio detected SFGs), we place an indicative upper limit to $L^{\rm AGN}_{\rm 1.4}$ (see left-pointing arrow in (panel {\it c}) of Fig. \ref{fig:merged}), consistent with the 5$\sigma$ luminosity limit of our 3~GHz survey. Each upper limit was centered at the mean redshift of the underlying population and scaled to 1.4~GHz by assuming a single power-law spectrum with spectral index $\alpha$=--0.7 (e.g. \citealt{Condon1992}). 

The f$_{\rm SF}$ estimates derived for this control sample are in good agreement with those calculated for radio-excess AGN within the same redshift bin. 
While it is well accepted that radio AGN activity is more prevalent in massive galaxies (\citealt{Fabian2012}; \citealt{Morganti+2013}), these findings (Fig. \ref{fig:merged}, panel {\it c}) suggest that, at fixed M$_{\star}$ and redshift, the triggering of radio AGN activity is independent of the color and star-forming properties of the host galaxy. These results broadly support the idea that the overall fraction of star-forming galaxies significantly increases with redshift, regardless of $L^{\rm AGN}_{\rm 1.4}$, which might explain the similar redshift dependence seen for the average s-BHAR in radio AGN.

\section{Discussion} \label{discussion}

We analysed a complete, well-defined sample of radio-excess AGN selected at 3~GHz in the COSMOS field. Stacking of deep X-ray images obtained with \textit{Chandra} was performed for 1272 radio AGN across a wide luminosity and redshift range, which allowed us to infer the average $L^{\rm AGN}_{\rm X}$ and s-BHAR in each bin. In this Section, we discuss and interpret the observed trends of s-BHAR as a function of $L^{\rm AGN}_{\rm 1.4}$ and redshift. The average estimates of the AGN radiative power obtained from X-ray stacking (Sect. \ref{average_sbhar}) display a positive evolution with redshift, yielding $\lambda_{\rm edd}$ values consistent with radiatively efficient SMBH accretion ($\lambda_{\rm edd} \gtrsim$1 per cent) at $z\gtrsim2$, at fixed 1.4~GHz luminosity $L^{\rm AGN}_{\rm 1.4}$. We also observed a qualitatively similar evolution in the fraction of star-forming host galaxies, possibly hinting at a connection between AGN radiative power and active star formation in radio AGN hosts. In order to investigate further what might be driving the observed trends, we discuss the main properties of radio-excess AGN hosts and compare our results with previous studies in the literature.

\subsection{Radio-excess AGN hosts through cosmic time} \label{rex_hosts}

The AGN radiative power traced by X-ray emission is commonly associated with on-going SMBH accretion (e.g. \citealt{Alexander+2012}). Several observational studies pointed out that the volume-averaged SMBH accretion density and SFR density evolve in a similar fashion, both peaking at redshift $z\sim$1--3 (e.g. \citealt{Madau+2014}), which thereby suggests a statistical co-evolution between AGN and galaxy growth. It is widely accepted that SMBH growth and galaxy star formation are both fuelled via cold gas accretion. Indeed, X-ray AGN, tracing the radiative phase of SMBH accretion, have been found to prefer star-forming, gas-rich galaxies (\citealt{Rosario+2013}; \citealt{Vito+2014}). Moreover, the intrinsic s-BHAR distribution of AGN was found to shift towards higher values with increasing redshift, at fixed M$_{\star}$, and especially in star-forming galaxies (e.g. \citealt{Aird+2018}). These findings support the idea of cold gas being the key ingredient for triggering AGN and star formation activity. 

In Sect. \ref{colours}, we examined the star-forming content of radio-excess AGN hosts by exploiting the $(NUV-r)$/$(r-J)$ colours as dust-insensitive tracers of active star formation. Previous studies investigating the properties of radio AGN hosts have relied on the presence of \textit{Herschel} detection in the far-infrared (FIR) to assess the star-forming content of the host (e.g. \citealt{Magliocchetti+2018}). However, such a selection is limited by the \textit{Herschel} sensitivity, that might be missing a substantial fraction of galaxies with relatively low SFRs, especially at high redshift. On the contrary, the $(NUV-r)$ / $(r-J)$ selection used in this work has been proven to be effective in selecting typical main-sequence galaxies also at $z \gtrsim 4$ (e.g. \citealt{Ilbert+2013}; \citealt{Davidzon+2017}), in the absence of strong AGN contamination. 

Fig. \ref{fig:merged} (panel {\it c}) displays a weak trend of f$_{\rm SF}$ with $L^{\rm AGN}_{\rm 1.4}$, at fixed redshift, while the dependence on redshift appears significantly stronger. However, this redshift trend could be mainly driven by the cosmic evolution of the full galaxy population (see Sect. \ref{colours}). Indeed, as mentioned in Sect. \ref{colours}, we observed a similar f$_{\rm SF}$ behaviour for a control sample of (redshift and M$_{\star}$-matched) galaxies without radio excess. This implies that the overall galaxy population follows a similar transformation through cosmic time, regardless of whether an AGN is driving strong radio jets or not. According to this scenario, the cosmic evolution of galaxies is mainly driven by the evolution of their cold gas content, which may occasionally trigger and sustain AGN activity. 

In fact, independent works found that star-forming galaxies around the ``main-sequence'' relation (MS, \citealt{Noeske+2007}; \citealt{Elbaz+2011}) become increasingly richer in cold gas (M$_{\rm gas}$) towards higher redshifts, with their molecular gas fraction (f$_{\rm gas}$=M$_{\rm gas}$/M$_{\star}$) steeply rising from $z\sim 0$ to $z\sim2-3$ (\citealt{Daddi+2010}; \citealt{Tacconi+2010}; \citealt{Geach+2011}; \citealt{Magdis+2012}; \citealt{Saintonge+2012}). Moreover, \citet{Gobat+2017} found a surprisingly large dust and gas content in quiescent galaxies at $z\sim$1.8, containing 2--3 orders of magnitude more dust, at fixed M$_{\star}$, than local quiescent galaxies. This indicates that early-type galaxies at $z>1$ are not truly passive, but instead contain substantial amounts of cold gas (5--10 per cent), which could possibly sustain (though in a smaller contribution compared to SFGs) radiatively-efficient SMBH accretion. 

All these studies further corroborate the idea that (i) the high-redshift Universe facilitates radiative AGN activity; and (ii) the evolution of the cold gas content shapes the cosmic transformation of galaxies.

In support of this interpretation, Fig. \ref{fig:merged} (panel {\it d}) shows that the average s-BHAR calculated for radio-excess AGN within ``blue'' (circles) star-forming hosts is systematically ($>$3$\times$) higher than that derived for the ``red'' (squares) quiescent subsample, based on the $(NUV-r)$ / $(r-J)$ criterion. This difference is seen out to $z\sim1.5$, while at $z\gtrsim2$ the red subsample is either absent or shows upper limits (down-pointing arrows) due to poor statistics.

\begin{figure}
\begin{center}
    \includegraphics[width=\linewidth]{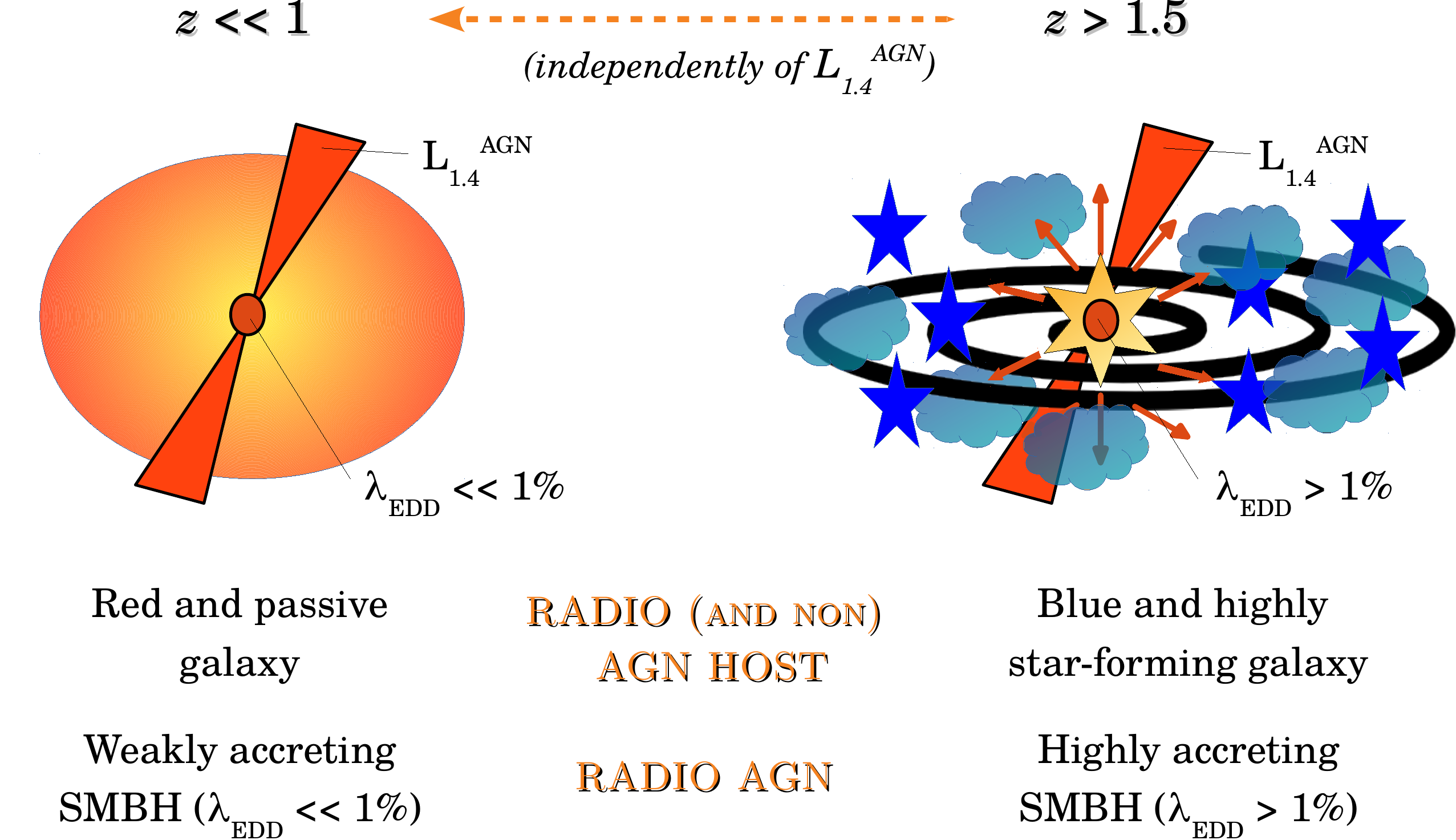}
\end{center}
 \caption{\small 
 Cartoon summarising a possible evolutionary scheme for radio-excess AGN of a given radio power, based on the results of this work.}
   \label{fig:cartoon}
\end{figure}

A simplistic cartoon shown in Fig. \ref{fig:cartoon} summarises this possible evolutionary scenario for radio AGN. Broadly, a radio AGN with a given $L^{\rm AGN}_{\rm 1.4}$ undergoes a strong evolution of the average radiative AGN power from low to high redshift, reaching $\lambda_{\rm edd} \gtrsim$1~per cent at $z\gtrsim$1.5, independently of $L^{\rm AGN}_{\rm 1.4}$. In parallel, the overall galaxy population evolves with redshift from red and quiescent to blue and star-forming systems, regardless of whether a galaxy is hosting a radio AGN or not.

These results argue in favour of a connection between AGN radiative power and star-forming content of radio AGN hosts. Testing this picture for the overall galaxy population would require a M$_{\star}$-selected galaxy sample, as well as stacking deep X-ray/radio images to assess the radiative and kinetic AGN emission in galaxies at different cosmic epochs. Future work from our team will undertake such an analysis.

\subsection{SMBH growth in radio-excess AGN} \label{smbh_rex_agn}

Several studies argued that radio-excess AGN are likely a mixture of radiatively efficient (``radiative mode'') and inefficient (``jet mode'') SMBHs, which have been found to evolve differently with cosmic time (\citealt{Willott+2001}; \citealt{Best+2014}; \citealt{Pracy+2016}; \citealt{Smolcic+2017c}; \citealt{Novak+2018}). In particular, while radiative-mode AGN roughly follow the evolution of the cosmic SFR density, consistently with them being fuelled via cold gas accretion, jet-mode AGN display a decrease in space density that resembles the decline of massive quiescent galaxies with redshift (\citealt{Best+2014}). 

The strong redshift dependence seen in the average s-BHAR, at fixed $L^{\rm AGN}_{\rm 1.4}$ (panel {\it b} of Fig. \ref{fig:merged} and Fig. \ref{fig:lr_vs_z_edd}) is probably linked to the stronger cosmic evolution seen for the radiative-mode AGN population. An AGN synthesis model describing the various states in which SMBH growth may occur was presented by \citet{Merloni+2008}. This model postulates three accretion modes, as a function of $\lambda_{\rm edd}$ and released power: (i) the ``low-kinetic'' (LK) mode, in which most of the power is released in kinetic form, occurs below $\lambda_{\rm edd} \sim$ few per cent; (ii) the ``high radiative'' (HR) mode, in which the power is mainly released in radiative form with $\lambda_{\rm edd} >$~few per cent; (iii) the ``high kinetic'' (HK) mode, in which radiative and kinetic power are both high ($\lambda_{\rm edd} >$~few per cent) and comparable. According to this model, in the radiatively-efficient regime, radio-selected AGN can trace both kinetic and radiative power.

In the light of these considerations, the lack of correlation between s-BHAR and $L^{\rm AGN}_{\rm 1.4}$ seen in Fig. \ref{fig:merged} (panel {\it b}), at fixed redshift, might be explained if synchrotron (radio) and accretion disk (X-ray) emission were statistically decoupled from one another, in a statistical sense. In addition, especially for the brightest radio sources, the different spatial scales at which AGN-driven X-ray (sub-pc) and radio ($>$kpc) emission processes operate could wash out possible hints of correlation.

On the other hand, the positive redshift dependence observed for a given $L^{\rm AGN}_{\rm 1.4}$ suggests that radio AGN become more radiatively efficient towards early cosmic epochs. If the cosmic evolution of galaxies' colours is driven by their f$_{\rm gas}$, this trend also justifies the systematic increase in the average s-BHAR with redshift, thus reaching SMBH accretion rates consistent with the so-called HK and HR modes ($\lambda_{\rm edd}>$ few per cent). Interestingly, our VLA-COSMOS data reveal that the average 3~GHz sizes of radio-excess AGN at $z\sim2$ appear to be significantly ($>$3$\sigma$) larger than those of their lower redshift ($z<$ 1) analogs, at fixed radio luminosity (Bondi et al., in prep.). Since radiatively inefficient jet-mode accretion is usually associated with a compact radio core, an increasing size evolution might be the signature of a possible change in the typical accretion mode of radio AGN. This idea is supported by observational studies finding that the intrinsic s-BHAR (or $\lambda_{\rm edd}$) distribution traced by X-ray emission shifts towards higher values with increasing redshift \citep{Aird+2018}, and especially in star-forming systems. Understanding whether the average radio AGN emission follows the same trend, would require an in-depth analysis of the overall $L^{\rm AGN}_{\rm X}$/$L^{\rm AGN}_{\rm 1.4}$ distribution, that will be presented in a separate paper (Delvecchio et al., in prep).

We note that our sample of radio-excess AGN achieves an unprecedented combination of completeness and sensitivity, and it does not exclusively consist of ``radio loud'' AGN (e.g. selected above a certain 1.4~GHz power). On the contrary, this sample spans the full 3~GHz flux density range down to F$_{\rm 3 GHz} \sim$~11~$\mu$Jy (i.e. 5$\sigma$ sensitivity), which allows us to draw conclusions that are valid for the overall radio AGN population over a wide luminosity and redshift range. However, we acknowledge that a fraction of radio AGN within the $L^{\rm AGN}_{\rm 1.4}$--$z$ space analysed in this work might be missed, if this AGN-driven radio emission is washed out by the star formation contribution in radio, thus failing to meet the required 2$\sigma$ cut in radio excess. We tested this possible incompleteness by relaxing the radio-excess threshold presented in Sect. \ref{agn_class} from 2$\sigma$ to 1$\sigma$. This increases the radio-excess sample (within the same $L^{\rm AGN}_{\rm 1.4}$--$z$ space analysed in this work) by 40 per cent, especially towards faint 1.4~GHz luminosities. Though this change introduces more star-forming hosts within our sample, we verified that the main trends with redshift and $L^{\rm AGN}_{\rm 1.4}$ persist, thus confirming our previous results. 
However, we estimate that 1/3 of the newly-selected (at $>$1$\sigma$ radio excess) AGN might be contaminated by SFGs, which motivates our original choice of considering the 2$\sigma$ radio-excess AGN sample.
A more complete and unbiased view of radio AGN activity could be reached by combining deep VLA and high-resolution VLBI data ($<$0.01 arcsec) to overcome the problem of galaxy dilution towards fainter radio fluxes (e.g. \citealt{HerreraRuiz+2017}).\\

This work provides a comprehensive view of the cosmic behaviour of radio AGN and their hosts. While previous studies, limited to radio-bright AGN ($L^{\rm AGN}_{\rm 1.4} \gtrsim$10$^{25}$~ W~Hz$^{-1}$) or intermediate redshifts ($z\lesssim1$), found radio AGN to reside within red and passive galaxies, our sample explores \textit{both} fainter sources and higher redshifts. This allowed us to unveil a heterogeneous population of radio AGN hosts, displaying a strong cosmic evolution in terms of optical colours, star formation and SMBH accretion rates. Though the mechanisms responsible for triggering radio AGN activity are still unconstrained, we argue that the presence of radio AGN does not seem to affect the star-forming content of galaxies, at any redshift and radio luminosity. On the other hand, our results imply that radio AGN, independently of their power, display a positive evolution in the AGN radiative power, gradually reaching a radiatively-efficient accretion mode at $z\gtrsim2$.

\section{Conclusions} \label{conclusions}

This work presents a comprehensive analysis of the average SMBH accretion properties of radio-excess AGN. Our sample was originally selected from the VLA-COSMOS~3~GHz~Large~Project \citep{Smolcic+2017a}, the deepest radio survey ever carried out in the COSMOS field. Our sample includes about 1800 radio-selected AGN, which were identified via a ($>$2$\sigma$) radio excess relative to the IRRC of star-forming galaxies in the COSMOS field \citep{Delhaize+2017}. To mitigate possible selection effects, we further selected a subset of 1272 radio AGN that is complete in $L^{\rm AGN}_{\rm 1.4}$, spanning the redshift range $0.6 <z< 5$ down to $L^{\rm AGN}_{\rm 1.4} \sim$10$^{23}$~ W~Hz$^{-1}$. This sample allowed us to explore the average SMBH accretion properties \textit{simultaneously} as a function of $L^{\rm AGN}_{\rm 1.4}$ and redshift. We stacked deep and uniform \textit{Chandra} images in 13 different bins of the $L^{\rm AGN}_{\rm 1.4}$--$z$ space, in order to infer the average X-ray emission and SMBH accretion properties of radio-excess AGN. We summarize our main conclusions as follows:
\begin{enumerate}
 \item The average $L^{\rm AGN}_{\rm X}$ of radio-excess AGN displays a positive evolution with redshift, at all $L^{\rm AGN}_{\rm 1.4}$, while we observe no correlation with $L^{\rm AGN}_{\rm 1.4}$, at fixed redshift (see panel {\it a} of Fig. \ref{fig:merged}).
 
 \item The average s-BHAR, tracing the AGN radiative power, increases by a factor of ten from $z\sim$0.6 to $z\sim$3.5, at fixed $L^{\rm AGN}_{\rm 1.4}$ (see panel {\it b} of Fig. \ref{fig:merged} and Fig. \ref{fig:lr_vs_z_edd}). If expressed in terms of Eddington ratio, our results suggest that SMBH accretion in radio AGN becomes radiatively efficient ($\lambda_{\rm edd} \gtrsim$1 per cent) at $z\gtrsim2$, differently from what seen at lower redshift ($z\lesssim1$) for radio-bright ($L^{\rm AGN}_{\rm 1.4} \gtrsim$10$^{25}$~ W~Hz$^{-1}$) AGN (e.g. \citealt{Hickox+2009}).
 
 \item The host galaxies of radio AGN display a strong evolution in their $(NUV-r)$/$(r-J)$ colours, becoming progressively bluer and more star-forming with redshift. The fraction of ``blue'' radio AGN hosts (from the criterion of \citealt{Davidzon+2017}) increases from 20 to $>$80 per cent across the redshift range 0.7 $<z<$ 3.5, roughly independently of the radio power (see panel {\it c} of Fig. \ref{fig:merged}). This redshift trend qualitatively resembles that observed for the average s-BHAR (or $\lambda_{\rm edd}$), possibly suggesting a statistical connection between star-forming content and radiative AGN activity (e.g. \citealt{Vito+2014}). These two trends are consistent with a scenario in which the molecular gas content within the host galaxy drives both SMBH accretion and star formation in radio AGN hosts.
 
 \item The strong evolution in the colours of radio AGN hosts is fully consistent with the average colours derived for a control sample of non-AGN galaxies, matched in redshift and M$_{\star}$ (see panel {\it c} of Fig. \ref{fig:merged}). This finding suggests that ``jet-mode'' feedback traced by radio AGN activity does not significantly affect the overall evolution of galaxies' colours. Whatever the mechanisms responsible for triggering radio AGN activity are, they seem to work independently of the star-forming content of the host.
 
\end{enumerate}

Our results challenge the (often assumed) association between radio AGN activity and red passive host, and we interpret our findings within a ``big picture'' of AGN-galaxy evolution. This work shows that the overall population of galaxies follows a strong redshift evolution in terms of colours and star-forming content, at least out to $z\sim2$. In this scenario, galaxies hosting radio AGN are no exception, as they display a consistent behaviour with a control sample of (M$_{\star}$--matched) non-AGN galaxies, at all redshifts. Our data suggest that jet-driven AGN activity does not primarily influence the cosmic transformation of galaxies. On the other hand, the increasing availability of cold gas supply (indicated by the $(NUV-r)$/$(r-J)$ colours) in the high-redshift Universe appears to facilitate radiative AGN activity. This corroborates the idea that radio-emitting AGN at high redshift ($z>1.5$) do not exclusively trace jet-mode feedback, bur rather a combination of both radiative and kinetic output, in which both accretion modes are effective and comparable. These clues might be useful for advancing our current understanding on how AGN feeback comes into play and how it regulates the evolution of galaxies at different cosmic epochs.

\section*{Acknowledgements}
The authors are grateful to the anonymous referee for his/her constructive comments that improved the content of this manuscript. ID is grateful to Fabio La Franca and Giorgio Lanzuisi for useful discussions. ID, VS, MN, JD acknowledge the European Union's Seventh Framework programme under grant agreement 337595 (ERC Starting Grant, ``CoSMass''). TM and {\sc CSTACK} are supported by UNAM\_DGAPA PAPIIT IN104216 and CONACyT 252531. 
DR and DMA acknowledge the support of the Science and Technology Facilities Council (STFC) through grant ST/P000541/1.



\bibliographystyle{mnras}
\bibliography{biblio} 




\bsp	

\label{lastpage}

\end{document}